\documentclass{article}%
\usepackage{amsmath}
\usepackage{graphicx}
\usepackage[mathscr]{eucal}%
\usepackage{amsfonts}%
\usepackage{amssymb}
\renewcommand{\thefootnote}{\arabic{footnote}}
\renewcommand{\mathit}{\mathscr}

\begin{document}

\title{Local Superfield Lagrangian BRST Quantization}
\author{\textsc{D.M.~Gitman}$^{a)}$\thanks{E-mail: gitman@dfn.if.usp.br},
\textsc{P.Yu. Moshin}$^{a),b)}$\thanks{E-mail: moshin@dfn.if.usp.br}, and
\textsc{A.A. Reshetnyak}$^{b)}$\thanks{E-mail: reshet@tspu.edu.ru}\\ \\$^{a)}$Instituto de F\'{\i}sica, Universidade de S\~{a}o Paulo,\\Caixa Postal 66318-CEP, 05315-970 S\~{a}o Paulo, S.P., Brazil\\$^{b)}$Tomsk State Pedagogical University, 634041 Tomsk, Russia}
\date{}
\maketitle

\begin{abstract}
A $\theta$-local formulation of superfield Lagrangian quantization in
non-Abelian hypergauges is proposed on the basis of an extension of general
reducible gauge theories to special superfield models with a Grassmann
parameter $\theta$. We solve the problem of describing the quantum action and
the gauge algebra of an $L$-stage-reducible superfield model in terms of a
BRST charge for a formal dynamical system with first-class constraints of
$(L+1)$-stage reducibility. Starting from $\theta$-local functions of the
quantum and gauge-fixing actions, with an essential use of Darboux coordinates
on the antisymplectic manifold, we construct superfield generating functionals
of Green's functions, including the effective action. We present two
superfield forms of BRST transformations, considered as $\theta$-shifts along
vector fields defined by Hamiltonian-like systems constructed in terms of the
quantum and gauge-fixing actions and an arbitrary $\theta$-local boson
function, as well as in terms of corresponding fermion functionals, through
Poisson brackets with opposite Grassmann parities. The gauge independence of
the S-matrix is proved. The Ward identities are derived. Connection is
established with the BV method, the multilevel Batalin--Tyutin formalism, as
well as with the superfield quantization scheme of Lavrov, Moshin, and
Reshetnyak, extended to the case of general coordinates.

\end{abstract}

\section{Introduction}

The construction of superfield counterparts of the Lagrangian \cite{BV} and
Hamiltonian \cite{BFV,Henneaux} quantization schemes for gauge theories on the
basis of BRST symmetry \cite{BRST} has been covered in a number of papers
\cite{BatalinBeringDamgaard1,LMR,GLM}. These works are based on nontrivial
(represented by the operator $D=\partial_{\theta}+\theta\partial_{t}$,
$[D,D]_{+}=2\partial_{t}$) and trivial relations between the even $t$ and odd
$\theta$ components of supertime\footnote{One of the first attempts to apply
the concept of odd time to Lagrangian quantization was made in \cite{Dayi}.}
\cite{Shander}. In \cite{BatalinBeringDamgaard1,LMR,GLM}, the geometric
interpretation \cite{BonoraToninBaulieu} of BRST transformations is realized
by special translations in superspace, which originally provided a basis for
the superspace description \cite{Hull} of quantum theories of Yang--Mills type.

It should be noted that superfield quantization is closely related to
generalized Poisson sigma-models, used in \cite{CattaneoFelder} to realize the
concept of star-product within the deformation quantization of Poisson
manifolds \cite{Kontsevich}, described from a superfield geometric viewpoint
in \cite{AlexandrovKontsevichSchwarzZaboronsky} and developed algorithmically
by Batalin and Marnelius in \cite{BatalinMarnelius1}. The geometry of $D=2$
supersymmetric sigma-models \cite{Hull1} with an arbitrary, $N\geq1$, number
of Grassmann coordinates was adapted to the classical and quantum description
of $D=1$ sigma-models by Hull, and, independently, the introduction of $N=2$
nilpotent parameters was applied to the construction of the partition function
for a classical mechanics by Gozzi et al \cite{Gozzi}. Quantization with a
single fermion supercharge, $Q(t,\theta)$, containing the BRST charge and the
unitarizing Hamiltonian \cite{BatalinBeringDamgaard1}, was recently extended
to $N=2$ (non-spacetime) supersymmetries \cite{BatalinDamgaard}, and then, in
\cite{BatalinBering}, to the case of an arbitrary number of supercharges,
$Q^{k}(t,\theta^{1},...,\theta^{N})$, $k=1,...,N$, depending on Grassmann
variables $\theta^{k}$. The superfield modification \cite{GrigorievDamgaard}
of the procedure \cite{BatalinBeringDamgaard1} reveals a close interplay
between the quantum action of the Batalin--Vilkovisky (BV) method \cite{BV}
and the BRST charge of the Batalin--Fradkin--Vilkovisky (BFV) method
\cite{BFV}. Finally, note that the superfield approach is used in the
description of second-class constrained systems as gauge models
\cite{BatalinMarnelius2} as well as in the second quantization of gauge
theories \cite{BarnichGrigoriev}.

The Lagrangian superfield partition function of \cite{BatalinBeringDamgaard1}
is derived from the Hamiltonian partition function through functional
integration over so-called Pfaffian ghosts and momenta. On the other hand, the
quantization rules \cite{LMR,GLM} present a superfield modification of the BV
formalism by including non-Abelian hypergauges \cite{BatalinTyutin}. The
corresponding hypergauge functions enter a gauge-fixing action which obeys
(following the ideas of \cite{BatalinBeringDamgaard2}) the same generating
equation that holds for the quantum action \cite{LMR,GLM}, except that the
first-order operator $V$ in this equation is replaced by the first-order
operator $U$ (these operators are crucial ingredients of \cite{LMR,GLM} from
the viewpoint of a superspace interpretation of BRST\ transformations).

The formalism \cite{LMR, GLM} provides a comparatively detailed analysis of
superfield quantization (BRST\ invariance, S-matrix gauge-independence). This
analysis is based on solutions to the generating equations; however, a
detailed correspondence between these solutions and a gauge model is not
established. To achieve a better understanding of the quantum properties based
on solutions of the superfield generating equations, it is natural to equip
the formalism \cite{LMR, GLM} with an \emph{explicit\ superfield description}
of gauge algebra structure functions that determine a given model. So far,
this problem has remained unsolved. For instance, the definition of a
classical action of superfields, ${\mathcal{A}}^{i}(\theta)=A^{i}+{\lambda
}^{i}\theta$, on a superspace with coordinates $(x^{\mu},\theta)$,
$\mu=0,\ldots,D-1$, as an integral of a nontrivial odd density, ${\mathcal{L}%
}(\mathcal{A}(x,\theta)$, $\partial_{\mu}{\mathcal{A}}(x,\theta),\ldots
;x,\theta)\equiv{\mathcal{L}}(x,\theta)$, is a problem for every given model.
Here, by trivial densities ${\mathcal{L}}(x,\theta)$ we understand those of
the form%
\[
\int d^{D}x\,d\theta\,{\mathcal{L}}(x,\theta)=\int d\theta\,\theta
\,S_{0}\left(  \mathcal{A}(\theta)\right)  =S_{0}(A),
\]
where $S_{0}(A)$\ is a usual classical action.

A peculiar feature of the vacuum functional $Z$ and generating functional of
Green's functions $Z[\Phi^{\ast}]$ in the formalism \cite{LMR,GLM} is the
dependence of the gauge fermion $\Psi\lbrack\Phi]$ and quantum action
${S}[\Phi,{\Phi}^{\ast}]$ on the components $\lambda^{A}$ of superfields
$\Phi^{A}(\theta)$ in the multiplet $(\Phi^{A},{\Phi}_{A}^{\ast}%
)(\theta)=(\phi^{A}+\lambda^{A}\theta,\phi_{A}^{\ast}-\theta J_{A})$, where
$(\phi^{A},\phi_{A}^{\ast},\lambda^{A},J_{A})$ are the complete set of
variables of the BV method. Another peculiarity of \cite{LMR,GLM} is that, due
to the manifest structure of $\Phi_{A}^{\ast}(\theta)$ and $Z[\Phi^{\ast}]$,
an effective action $\Gamma$ with the standard Ward identity $(\Gamma
,\Gamma)=0$ in terms of a superantibracket \cite{LMR} can be introduced by a
Legendre transformation of $\ln Z[\Phi^{\ast}]$ with respect to $P_{1}%
(\theta)\Phi_{A}^{\ast}(\theta)$,\footnote{The objects $P_{1}(\theta)$ and
$\delta/\delta\left(  P_{1}(\theta)\Phi_{A}^{\ast}(\theta)\right)  $ are,
respectively, an element of the system of projectors $\{P_{a}(\theta
)=\delta_{a0}(1-\theta\partial_{\theta})+\delta_{a1}\theta\partial_{\theta
},a=0,1\}$, acting on the supermanifold with coordinates $(\Phi^{A},{\Phi}%
_{A}^{\ast})(\theta)$, and a superfield variational derivative with respect to
$P_{1}(\theta)\Phi_{A}^{\ast}(\theta)$.}%
\begin{equation}
\Gamma\lbrack P_{0}(\Phi,\Phi^{\ast})]=\frac{\hbar}{i}\ln Z[\Phi^{\ast
}]+\partial_{\theta}\left\{  \left[  P_{1}(\theta)\Phi_{A}^{\ast}%
(\theta)\right]  \Phi^{A}(\theta)\right\}  ,\;\Phi^{A}(\theta)=-\frac{\hbar
}{i}\frac{\delta\ln Z[\Phi^{\ast}]}{\delta\left(  P_{1}(\theta)\Phi_{A}^{\ast
}(\theta)\right)  }\,. \label{1}%
\end{equation}
Although non-contradictory, such an introduction of $\Gamma$ violates the
superfield content of the variables.\footnote{By violating the superfield
content, we understand the fact that the derivative of $Z[\Phi^{\ast}]$, which
defines the effective action through a Legendre transformation, is taken with
respect to only one superfield component, namely, the $\theta$-component of
$\Phi_{A}^{\ast}\left(  \theta\right)  $, so that the resulting effective
action depends only on $\phi^{A}$ and $\phi_{A}^{\ast}$, which can be formally
expressed as $P_{0}\left(  \theta\right)  \left(  \Phi^{A},\Phi_{A}^{\ast
}\right)  \left(  \theta\right)  =\left(  \phi^{A},\phi_{A}^{\ast}\right)  $.}

In this paper, we propose a local formalism of superfield Lagrangian
quantization in which the quantities of an initial classical theory are
realized in the framework of a $\theta$\emph{-local superfield model} (LSM).
The idea of LSM is to represent the objects of a gauge theory (classical
action, generators of gauge transformations, etc.) in terms of $\theta$-local
functions, trivially{\ related to the spacetime coordinates, in the sense that
(as compared to the formalism \cite{BatalinBeringDamgaard1}) the derivatives
with respect to the even $t$ and odd $\theta$ component of supertime are taken
independently. Using an analogy with classical mechanics (or classical field
theory), we reproduce the dynamics and gauge invariance (in particular, BRST
transformations) of the initial theory (the one with $\theta=0$) in terms of
$\theta$-local equations, called \emph{Lagrangian }and\emph{\ Hamiltonian
systems} (LS, HS) with a dynamical odd time $\theta$, which implies that this
coordinate enters an LS or HS not as a parameter but as part of a differential
operator $\partial_{\theta}$ that describes the $\theta$-evolution of a
system. }

On the basis of the suggested formalism, we circumvent the peculiarities of
the functionals $Z$ and $Z[\Phi^{\ast}]$ in \cite{LMR,GLM} as well as solve
the following problems:

1. We develop a \emph{dual description} of an arbitrary reducible LSM of Ref.
\cite{GrigorievDamgaard} in the case of irreducible gauge theories (with
bosonic classical fields and parameters of gauge transformations), in terms of
a BRST charge related to a (formal) dynamical system with first-class
constraints of a higher stage of reducibility. By dual\emph{\ }description, we
understand such a treatment of a gauge model that interrelates the Lagrangian
and Hamiltonian formulations (the latter understood in the sense of
formal\emph{\ }dynamical systems).

2. An HS constructed from $\theta$-local quantities, i.e., a quantum action, a
gauge-fixing action, and an arbitrary bosonic function, encodes (through a
$\theta$-local antibracket) both BRST and so-called anticanonical-like
transformations, in terms of a universal set of equations underlying the
gauge-independence of the S-matrix. This set of equations is generated, in
terms of an even superfield Poisson bracket, by a linear combination of
fermionic functionals corresponding to the mentioned $\theta$-local
quantities, e.g., the quantum and gauge-fixing actions as well as the bosonic function.

3. For the first time in the framework of superfield approach, we introduce a
\emph{superfield effective action} (also in the case of non-Abelian hypergauges).

4. We extend the superfield quantization of Refs. \cite{LMR,GLM} to the case
of general coordinates on the manifold of super(anti)fields and establish a
relation with the proposed local quantization.

The paper is organized as follows. In Section 2, a Lagrangian formulation of
an LSM is proposed as an extension of a usual model of classical fields
$A^{i}$, $i=1,...,n=n_{+}+n_{-}$, to a $\theta$-local theory, defined on the
odd tangent bundle $T_{\mathrm{odd}}\mathcal{M}_{\mathrm{CL}}\equiv\Pi
T\mathcal{M}_{\mathrm{CL}}$ = $\left\{  \mathcal{A}^{I},\partial_{\theta
}\mathcal{A}^{I}\right\}  $, $I=1,\ldots,N=N_{+}+N_{-}$\footnote{$\Pi$ denotes
the operation that changes the coordinates of a tangent fiber bundle
$T\mathcal{M}_{\mathrm{CL}}$ over a configuration $\mathcal{A}^{I}$ into the
coordinates of the opposite Grassmann parity \cite{Schwarz}, and $N_{+}$,
$N_{-}$ are the numbers of bosonic and fermionic fields, among which there may
exist superfields corresponding to the ghosts of the minimal sector in the BV
quantization scheme (in terms of the condensed notation \cite{DeWitt} used in
this paper).}, $(n_{+},n_{-})\leq(N_{+},N_{-})$. The superfields
$(\mathcal{A}^{I},\partial_{\theta}\mathcal{A}^{I})$ are defined in a
superspace $\mathcal{M}=\widetilde{\mathcal{M}}\times\widetilde{P}$
parameterized by $\left(  z^{M},\theta\right)  $, where the spacetime
coordinates $z^{M}\subset i\subset I$ include Lorentz vectors and spinors of
the superspace $\widetilde{\mathcal{M}}$. We investigate the superfield
equations of motion, introduce the notions of reducible \emph{general} and
\emph{special}\textit{\ }superfield gauge theories and apply Noether's first
theorem to $\theta$-translations. Section 3 is devoted to the Hamiltonian
formulation of an LSM on the odd cotangent bundle $T_{\mathrm{odd}}^{\ast
}\mathcal{M}_{\mathrm{CL}}\equiv\Pi T^{\ast}\mathcal{M}_{\mathrm{CL}}=\left\{
\mathcal{A}^{I},\mathcal{A}_{I}^{\ast}\right\}  $. Here, we establish a
connection to the Lagrangian formalism and investigate the existence of a
Noether integral, related to $\theta$-translations, that leads to the validity
of a $\theta$-local master equation. The quantization rules are given in
Section 4. In particular, we construct the dual description of an LSM and
define a generating functional of Green's functions, $\mathsf{Z}(\theta)$, and
an effective action, $\mathsf{\Gamma}(\theta)$, using an invariant description
of super(anti)fields on a general antisymplectic manifold. An essential
feature in introducing $\mathsf{Z}(\theta)$ and $\mathsf{\Gamma}(\theta)$ is a
choice of Darboux coordinates $(\varphi,\varphi^{\ast})(\theta)$ compatible
with the properties of the quantum action. In Section 5, on the basis of a
component form of the local superfield quantization, we establish its
connection with the first-level formalism \cite{BatalinTyutin}, with the BV
method, and with an extension of the superfield scheme \cite{LMR,GLM}. In the
Conclusion, we discuss the results of the present work.

In addition to DeWitt's condensed notation \cite{DeWitt}, we partially use the
conventions of Refs. \cite{LMR,GLM}. We distinguish between two types of
superfield derivatives: the right (left) variational derivative $\delta
_{(l)}F/\delta\Phi^{A}(\theta)$ of a functional $F$, and the right (left)
derivative $\partial_{(l)}\mathcal{F}(\theta)/\partial\Phi^{A}(\theta)$ of a
function $\mathcal{F}(\theta)$ for a fixed $\theta$. Derivatives with respect
to super(anti)fields and their components are understood as right (left), for
instance, $\delta/\delta\Phi_{A}^{\ast}(\theta)$ or $\delta/\delta\lambda^{A}%
$, while the corresponding left (right) derivatives are labelled by the
subscript $l(r)$. For right-hand derivatives with respect to $\mathcal{A}%
^{I}(\theta)$, with a fixed $\theta$, we use the notation $\mathcal{F}%
,_{I}(\theta)\equiv\partial\mathcal{F}(\theta)/\partial\mathcal{A}^{I}%
(\theta)$. The $\delta(\theta)$-function and integration over $\theta$ are
given, respectively, by $\delta(\theta)=\theta$ and left-hand differentiation
over $\theta$. We refer to a function $\mathcal{F}(\theta)$, regarded as an
element of the superalgebra $C^{\infty}(T_{\mathrm{odd}}\mathcal{M}%
_{\mathrm{CL}})$, as a $C^{\infty}(T_{\mathrm{odd}}\mathcal{M}_{\mathrm{CL}}%
)$-function. The rank of an even $\theta$-local supermatrix $K(\theta)$ with
$Z_{2}$-grading $\varepsilon$ is characterized by a pair of numbers
$\overline{m}=(m_{+},m_{-})$, where $m_{+}$ ($m_{-}$) is the rank of the
Bose--Bose (Fermi--Fermi) block of the $\theta$-independent part of the
supermatrix $K(\theta)$: $\mathrm{rank}\Vert K(\theta)\Vert=\mathrm{rank}\Vert
K(0)\Vert$. With respect to the same Grassmann parity $\varepsilon$, we
understand the dimension of a smooth supersurface, also characterized by a
pair of numbers, in the sense of the definition \cite{Berezin} of a
supermanifold, so that the above pair coincides with the corresponding numbers
of the Bose and Fermi components of $z^{i}(0)$, being the $\theta$-independent
parts of local coordinates $z^{i}(\theta)$ parameterizing this
supersurface\footnote{In the infinite-dimensional case (which we preferably
use in this paper) the concept of dimension has to be clarified. Thus, for the
vector bundle $\mathcal{M}_{\mathrm{CL}}\rightarrow\widetilde{\mathcal{M}}$,
we formally understand that $\dim\mathcal{M}_{\mathrm{CL}}$ is the dimension
of a fiber $\mathcal{F}_{p}^{\mathcal{M}_{\mathrm{CL}}}$ over an arbitrary
point $p\in\widetilde{\mathcal{M}}$.}. On these pairs, we consider the
operations of component addition, $\overline{m}+\overline{n}=(m_{+}%
+n_{+},m_{-}+n_{-})$, and comparison,%
\[
\overline{m}=\overline{n}\Leftrightarrow m_{\pm}=n_{\pm},\;\overline
{m}>\overline{n}\Leftrightarrow(m_{+}>n_{+},\;m_{-}\geq n_{-})\;\mathrm{or}%
\;(m_{+}\geq n_{-},\;m_{+}>n_{-}).
\]

\section{Odd-time Lagrangian Formulation}

The basic objects of the Lagrangian formulation of an LSM are a
\emph{Lagrangian action} $S_{\mathrm{L}}$: $T_{\mathrm{odd}}\mathcal{M}%
_{\mathrm{CL}}\times\{\theta\}$ $\rightarrow$ $\Lambda_{1}(\theta;\mathbb{R}%
)$, being a $C^{\infty}(T_{\mathrm{odd}}\mathcal{M}_{\mathrm{CL}})$-function
taking values in a real Grassmann algebra $\Lambda_{1}(\theta;\mathbb{R})$,
and (independently) a functional $Z[\mathcal{A}]$, whose $\theta$-density is
defined with accuracy up to an arbitrary function $f((\mathcal{A}%
,\partial_{\theta}\mathcal{A})(\theta),\theta)\in{\ker}\{\partial_{\theta}\}$,
$\vec{\varepsilon}(f)=\vec{0}$,%
\begin{equation}
Z[\mathcal{A}]=\partial_{\theta}S_{\mathrm{L}}(\theta),\;\vec{\varepsilon
}(Z)=\vec{\varepsilon}(\theta)=(1,0,1),\;\vec{\varepsilon}(S_{\mathrm{L}%
})=\vec{0}. \label{2}%
\end{equation}
The values $\vec{\varepsilon}=(\varepsilon_{P},\varepsilon_{\bar{J}%
},\varepsilon)$, $\varepsilon=\varepsilon_{P}+\varepsilon_{\bar{J}}$, of
$Z_{2}$-grading, with the auxiliary components $\varepsilon_{\bar{J}}$,
$\varepsilon_{P}$ related to the respective coordinates $\left(  z^{M}%
,\theta\right)  $ of a superspace $\mathcal{M}$, are defined on superfields
$\mathcal{A}^{I}(\theta)$ by the relation $\vec{\varepsilon}(\mathcal{A}%
^{I})=\left(  (\varepsilon_{P})_{I},(\varepsilon_{\bar{J}})_{I},\varepsilon
_{I}\right)  $. Note that $\mathcal{M}$ may be realized as the quotient of a
symmetry supergroup ${J}=\bar{J}\times P$, $P=\exp(i\mu p_{\theta})$, for the
functional $Z[\mathcal{A}]$, where $\mu$ and $p_{\theta}$ are, respectively, a
nilpotent parameter and a generator of $\theta$-shifts, whereas $\bar{J}$ is
chosen as the spacetime SUSY group. The quantities $\varepsilon_{\bar{J}}$,
$\varepsilon_{P}$ are the respective Grassmann parities of the coordinates of
representation spaces of the supergroups $\bar{J}$, $P$. The introduced
objects allow one to achieve a correct incorporation of the spin-statistic
relation into operator quantization.

Among the objects $S_{\mathrm{L}}(\theta)$ and $Z[\mathcal{A}]$, invariant
under the action of a $J$-superfield representation $T$ restricted to $\bar
{J}$, $\left.  T\right|  _{\bar{J}}$, it is only $S_{\mathrm{L}}(\theta)$ that
transforms nontrivially (in view of the $J$-scalar nature of $Z[\mathcal{A}]$)
with respect to the total representation $T$ under $\mathcal{A}^{I}%
(\theta)\rightarrow\mathcal{A}^{\prime}{}^{I}(\theta)=(\left.  T\right|
_{\bar{J}}\mathcal{A})^{I}(\theta-\mu)$,%
\begin{equation}
\delta S_{\mathrm{L}}(\theta)=S_{\mathrm{L}}\left(  \mathcal{A}^{\prime
}(\theta),\partial_{\theta}\mathcal{A}^{\prime}(\theta),{\theta}\right)
-S_{\mathrm{L}}(\theta)=-\mu\left[  \frac{\partial}{\partial\theta}%
+P_{0}(\theta)(\partial_{\theta}U)(\theta)\right]  S_{\mathrm{L}}(\theta).
\label{3}%
\end{equation}
Here, we have introduced the nilpotent operator $(\partial_{\theta}{U}%
)(\theta)=\partial_{\theta}\mathcal{A}^{I}(\theta)\partial_{l}/\partial
\mathcal{A}^{I}(\theta)=[\partial_{\theta},U(\theta)]_{-}$, $U(\theta
)=P_{1}\mathcal{A}^{I}(\theta)\partial_{l}/\partial\mathcal{A}^{I}(\theta)$.

Assuming the existence of a critical superfield configuration for
$Z[\mathcal{A}]$, one presents the dynamics of an LSM in terms of superfield
Euler--Lagrange equations:
\begin{equation}
\frac{\delta_{l}Z[\mathcal{A}]}{\delta\mathcal{A}^{I}(\theta)}=\left[
\frac{\partial_{l}}{\partial\mathcal{A}^{I}(\theta)}-(-1)^{\varepsilon_{I}%
}\partial_{\theta}\frac{\partial_{l}}{\partial(\partial_{\theta}\mathcal{A}%
{}^{I}(\theta))}\right]  S_{\mathrm{L}}(\theta)\equiv\mathcal{L}_{I}%
^{l}(\theta)S_{\mathrm{L}}(\theta)=0, \label{4}%
\end{equation}
equivalent, in view of $\partial_{\theta}^{2}\mathcal{A}^{I}(\theta)\equiv0$,
to an LS characterized by $2N$ formally second-order differential equations in
$\theta$,%
\begin{align}
&  \partial_{\theta}^{2}\mathcal{A}^{J}(\theta)\frac{\partial_{l}%
^{2}S_{\mathrm{L}}(\theta)}{\partial(\partial_{\theta}\mathcal{A}^{I}%
(\theta))\partial(\partial_{\theta}\mathcal{A}^{J}(\theta))}\equiv
\partial_{\theta}^{2}\mathcal{A}^{J}(\theta)(S_{\mathrm{L}}^{\prime\prime
})_{IJ}(\theta)=0,\nonumber\\
&  {\Theta}_{I}(\theta)\equiv\frac{\partial_{l}S_{\mathrm{L}}(\theta
)}{\partial\mathcal{A}^{I}(\theta)}-(-1)^{\varepsilon_{I}}\left[
\frac{\partial}{\partial\theta}\frac{\partial_{l}S_{\mathrm{L}}(\theta
)}{\partial(\partial_{\theta}\mathcal{A}^{I}(\theta))}+(\partial_{\theta
}U)(\theta)\frac{\partial_{l}S_{\mathrm{L}}(\theta)}{\partial(\partial
_{\theta}\mathcal{A}^{I}(\theta))}\right]  =0. \label{5}%
\end{align}
The \emph{Lagrangian constraints}{\thinspace}${\Theta}_{I}(\theta)={\Theta
}_{I}(\mathcal{A}(\theta),\partial_{\theta}\mathcal{A}(\theta),\theta
)${\thinspace}restrict{\thinspace}the setting of the{\thinspace}Cauchy problem
for the LS and may be functionally dependent, as first-order{\thinspace
equations in\thinspace}$\theta$.

Provided that there exists (at least locally) a supersurface $\Sigma
\subset\mathcal{M}_{\mathrm{CL}}$ such that%
\begin{equation}
\left.  {\Theta}_{I}(\theta)\right|  _{\Sigma}=0,\;\dim\Sigma=\overline
{M},\;\mathrm{rank}\left\|  \mathcal{L}_{J}^{l}(\theta)\left[  \mathcal{L}%
_{I}^{l}(\theta)S_{\mathrm{L}}(\theta)(-1)^{\varepsilon_{I}}\right]  \right\|
_{\Sigma}=\overline{N}-\overline{M}, \label{6}%
\end{equation}
there exist $M=M_{+}+M_{-}$ independent identities:%
\begin{equation}
\int d\theta\frac{\delta Z[\mathcal{A}]}{\delta\mathcal{A}^{I}(\theta)}%
{\hat{\mathcal{R}}}_{\mathcal{A}_{0}}^{I}(\theta;{\theta}_{0})=0,\;{\hat
{\mathcal{R}}}_{\mathcal{A}_{0}}^{I}(\theta;{\theta}_{0})=\sum_{k\geq0}\left(
\left(  \partial_{\theta}\right)  ^{k}\delta(\theta-\theta_{0})\right)
{\hat{\mathcal{R}}}_{k}{}_{\mathcal{A}_{0}}^{I}\left(  \mathcal{A}%
(\theta),\partial_{\theta}\mathcal{A}(\theta),\theta\right)  . \label{7}%
\end{equation}
The generators ${\hat{\mathcal{R}}}_{\mathcal{A}_{0}}^{I}(\theta;{\theta}%
_{0})$ of \emph{general gauge transformations},%
\[
\delta_{g}\mathcal{A}^{I}(\theta)=\int d\theta_{0}{\hat{\mathcal{R}}%
}_{\mathcal{A}_{0}}^{I}(\theta;{\theta}_{0}){\xi}^{\mathcal{A}_{0}}(\theta
_{0}),\;\vec{\varepsilon}({\xi}^{\mathcal{A}_{0}})=\vec{\varepsilon
}_{\mathcal{A}_{0}},\;\mathcal{A}_{0}=1,...,\;M_{0}=M_{0+}+M_{0-},
\]
that leave $Z[\mathcal{A}]$ invariant, are functionally dependent under the
assumption of locality and $\bar{J}$-covariance, provided that%
\[
\mathrm{rank}\left\|  \sum_{k\geq0}{\hat{\mathcal{R}}}_{k}{}_{\mathcal{A}_{0}%
}^{I}(\theta)\left(  \partial_{\theta}\right)  ^{k}\right\|  _{\Sigma
}=\overline{M}<\overline{M}_{0}.
\]
The dependence of ${\hat{\mathcal{R}}}_{\mathcal{A}_{0}}^{I}(\theta;{\theta
}_{0})$ implies the existence (on solutions of the LS) of proper
zero-eigenvalue eigenvectors, ${\hat{\mathcal{Z}}}{}_{\mathcal{A}_{1}%
}^{\mathcal{A}_{0}}\left(  \mathcal{A}(\theta_{0}),\partial_{\theta_{0}%
}\mathcal{A}(\theta_{0}),\theta_{0};\theta_{1}\right)  $, with a structure
analogous to ${\hat{\mathcal{R}}}_{\mathcal{A}_{0}}^{I}(\theta;{\theta}_{0})$
in (\ref{7}), which exhaust the zero-modes of the generators, and are
dependent in case%
\[
\mathrm{rank}\left\|  \sum_{k}\hat{\mathcal{Z}}_{k}{}_{\mathcal{A}_{1}%
}^{\mathcal{A}_{0}}(\theta_{0})\left(  \partial_{\theta_{0}}\right)
^{k}\right\|  _{\Sigma}=\overline{M}_{0}-\overline{M}<\overline{M}_{1}.
\]
As a result, the relations of dependence for eigenvectors that define a
general $L_{g}$-stage reducible LSM are given by%
\begin{align}
&  \hspace{-1em}\int d\theta^{\prime}{\hat{\mathcal{Z}}}_{\mathcal{A}_{s-1}%
}^{\mathcal{A}_{s-2}}(\theta_{s-2};{\theta}^{\prime}){\hat{\mathcal{Z}}%
}_{\mathcal{A}_{s}}^{\mathcal{A}_{s-1}}(\theta^{\prime};{\theta}_{s})=\int
d\theta^{\prime}{\Theta}_{J}(\theta^{\prime})\mathcal{L}_{\mathcal{A}_{s}%
}^{\mathcal{A}_{s-2}J}\left(  (\mathcal{A},\partial_{\theta}\mathcal{A}%
)(\theta_{s-2}),\theta_{s-2},\theta^{\prime};\theta_{s}\right)  ,\nonumber\\
&  \hspace{-1em}\overline{M}_{s-1}>\sum\limits_{k=0}^{s-1}(-1)^{k}\overline
{M}_{s-k-2}=\mathrm{rank}\left\|  \sum_{k\geq0}\hat{\mathcal{Z}}_{k}%
{}_{\mathcal{A}_{s-1}}^{\mathcal{A}_{s-2}}(\theta_{s-2})\left(  \partial
_{\theta_{s-2}}\right)  ^{k}\right\|  _{\Sigma},\nonumber\\
&  \hspace{-1em}\overline{M}_{L_{g}}=\sum\limits_{k=0}^{L_{g}}(-1)^{k}%
\overline{M}_{L_{g}-k-1}=\mathrm{rank}\left\|  \sum_{k\geq0}\hat{\mathcal{Z}%
}_{k}{}_{\mathcal{A}_{L_{g}}}^{\mathcal{A}_{L_{g}-1}}(\theta_{L_{g}-1})\left(
\partial_{\theta_{L_{g}-1}}\right)  ^{k}\right\|  _{\Sigma},\nonumber\\
&  \hspace{-1em}\vec{\varepsilon}({\hat{\mathcal{Z}}}_{\mathcal{A}_{s+1}%
}^{\mathcal{A}_{s}})=\vec{\varepsilon}_{\mathcal{A}_{s}}+\vec{\varepsilon
}_{\mathcal{A}_{s+1}}+(1,0,1),\ {\hat{\mathcal{Z}}}_{\mathcal{A}_{0}%
}^{\mathcal{A}_{-1}}(\theta_{-1};{\theta}_{0})\equiv{\hat{\mathcal{R}}%
}_{\mathcal{A}_{0}}^{I}(\theta_{-1};{\theta}_{0}),\nonumber\\
&  \hspace{-1em}\mathcal{L}_{\mathcal{A}_{1}}^{\mathcal{A}_{-1}J}(\theta
_{-1},\theta^{\prime};\theta_{1})\equiv\mathcal{K}_{\mathcal{A}_{1}}%
^{IJ}(\theta_{-1},\theta^{\prime};\theta_{1})=-(-1)^{(\varepsilon
_{I}+1)(\varepsilon_{J}+1)}\mathcal{K}_{\mathcal{A}_{1}}^{JI}(\theta^{\prime
},\theta_{-1};\theta_{1}). \label{8}%
\end{align}
for $s=1,...,L_{g}$, $\mathcal{A}_{s}=1,...$, $M_{s}=M_{s+}+M_{s-}$,
$\overline{M}\equiv\overline{M}_{-1}$. For $L_{g}=0$, the LSM is an
irreducible \emph{general gauge theory}.

In case an LSM admits the form $S_{\mathrm{L}}(\theta)=T\left(  \partial
_{\theta}\mathcal{A}(\theta)\right)  -S\left(  \mathcal{A}(\theta
),\theta\right)  $, the functions ${\Theta}_{I}(\theta)$ are given in the
extended configuration space $\mathcal{M}_{\mathrm{CL}}\times\{\theta\}$ by
the relations%
\begin{equation}
{\Theta}_{I}(\theta)=-S,_{I}\left(  {\mathcal{A}}(\theta),\theta\right)
(-1)^{\varepsilon_{I}}=0, \label{9}%
\end{equation}
being the extremals of the functional $S_{0}(A)=S\left(  \mathcal{A}%
(0),0\right)  $, corresponding to $\theta=0$. Condition (\ref{6}) and
identities (\ref{7}) take the usual form (in case $\theta=0$)
\begin{equation}
\mathrm{rank}\left\|  S,_{IJ}\left(  {\mathcal{A}}(\theta),\theta\right)
\right\|  _{\Sigma}=\overline{N}-\overline{M},\ \ S,_{I}\left(  \mathcal{A}%
(\theta),\theta\right)  {\mathcal{R}}_{0}{}_{\mathcal{A}_{0}}^{I}\left(
\mathcal{A}(\theta),\theta\right)  =0, \label{10}%
\end{equation}
with linearly-dependent (for $\overline{M}_{0}>\overline{M}$) generators of
\emph{special gauge transformations},%
\[
\delta\mathcal{A}^{I}(\theta)=\mathcal{R}_{0}{}_{\mathcal{A}_{0}}^{I}\left(
\mathcal{A}(\theta),\theta\right)  {\xi}_{0}^{\mathcal{A}_{0}}(\theta),
\]
that leave invariant only $S(\theta)$, in contrast to $T(\theta)$. The
dependence of generators $\mathcal{R}_{0}{}_{\mathcal{A}_{0}}^{I}(\theta)$, as
well as of their zero-eigenvalue eigenvectors $\mathcal{Z}_{\mathcal{A}_{1}%
}^{\mathcal{A}_{0}}(\mathcal{A}(\theta),\theta)$, and so on, can also be
expressed by special relations of reducibility for $s=1,...,L_{g}$, namely,%
\begin{align}
&  \mathcal{Z}_{\mathcal{A}_{s-1}}^{\mathcal{A}_{s-2}}(\mathcal{A}%
(\theta),\theta)\mathcal{Z}_{\mathcal{A}_{s}}^{\mathcal{A}_{s-1}}%
(\mathcal{A}(\theta),\theta)=S,_{J}(\theta)\mathcal{L}_{\mathcal{A}_{s}%
}^{\mathcal{A}_{s-2}J}(\mathcal{A}(\theta),\theta),\;\vec{\varepsilon
}(\mathcal{Z}_{\mathcal{A}_{s}}^{\mathcal{A}_{s-1}})=\vec{\varepsilon
}_{\mathcal{A}_{s-1}}+\vec{\varepsilon}_{\mathcal{A}_{s}},\nonumber\\
&  \mathcal{Z}_{\mathcal{A}_{0}}^{\mathcal{A}_{-1}}(\theta)\equiv
\mathcal{R}_{0}{}_{\mathcal{A}_{0}}^{I}(\theta),\;\mathcal{L}_{\mathcal{A}%
_{1}}^{\mathcal{A}_{-1}J}(\theta)\equiv\mathcal{K}_{\mathcal{A}_{1}}%
^{IJ}(\theta)=-(-1)^{\varepsilon_{I}\varepsilon_{J}}\mathcal{K}_{\mathcal{A}%
_{1}}^{JI}(\theta). \label{11}%
\end{align}

For $\overline{M}_{L_{g}}=\sum_{k=0}^{L_{g}}(-1)^{k}\overline{M}_{L_{g}%
-k-1}=\mathrm{rank}\left\|  \mathcal{Z}{}_{\mathcal{A}_{L_{g}}}^{\mathcal{A}%
_{L_{g}-1}}\right\|  _{\Sigma}$, relations (\ref{9})--(\ref{11}) determine a
\emph{special gauge theory} of $L_{g}$-stage reducibility. The gauge algebra
of such a theory is $\theta$-locally embedded into the gauge algebra of a
general gauge theory with the functional $Z[\mathcal{A}]=\partial_{\theta
}(T(\theta)-S(\theta))$, which implies the relation between the eigenvectors%
\begin{equation}
\hat{\mathcal{Z}}_{\mathcal{A}_{s}}^{\mathcal{A}_{s-1}}(\mathcal{A}%
(\theta_{s-1})\,,\theta_{s-1};\theta_{s})=-\delta(\theta_{s-1}-\theta
_{s})\mathcal{Z}_{\mathcal{A}_{s}}^{\mathcal{A}_{s-1}}(\mathcal{A}%
(\theta_{s-1}),\theta_{s-1}) \label{12}%
\end{equation}
and the fact that the structure functions of the gauge algebra of a special
gauge theory may depend on $\partial_{\theta}\mathcal{A}^{I}\left(
\theta\right)  $ only parametrically. Note that an extended (as compared to
$\{P_{a}\left(  \theta\right)  \}$, $a=0,1$) system of projectors onto
$C^{\infty}(T_{\mathrm{odd}}\mathcal{M}_{\mathrm{CL}})$ $\times$ $\{\theta\}$,
$\{{P}_{0}(\theta),\theta\partial/\partial\theta,U(\theta)\}$, selects from
(\ref{11}) two kinds of gauge algebras: one with structure equations and
functions $S(\mathcal{A}(\theta))$, $\mathcal{Z}_{\mathcal{A}_{s}%
}^{\mathcal{A}_{s-1}}(\mathcal{A}(\theta))$ not depending on $\theta$ in an
explicit form; another with the standard relations for the gauge algebra of a
reducible model with quantities $S_{0}(A)$, $\mathcal{Z}_{\alpha_{s}}%
^{\alpha_{s-1}}(A)$, in case $\theta=0$, $(\varepsilon_{P})_{I}=(\varepsilon
_{P})_{\mathcal{A}_{s}}=0$, $s=1,...,L_{g} $, and under the assumption of
completeness of the reduced generators $\mathcal{R}_{\alpha_{0}}%
^{i}(\mathcal{A}(\theta))$ and eigenvectors $\mathcal{Z}_{\alpha_{s}}%
^{\alpha_{s-1}}(\mathcal{A}(\theta))$; see Subsection 4.1.

An extension of a usual field theory to a $\theta$-local LSM permits one to
apply Noether's first theorem \cite{Noether} to the invariance of the density
$d\theta S_{\mathrm{L}}(\theta)$ with respect to global $\theta$-translations
as symmetry transformations of the superfields $\mathcal{A}^{I}(\theta)$ and
coordinates $(z^{M},\theta)$, $(\mathcal{A}^{I},z^{M},\theta)\rightarrow
(\mathcal{A}^{I},z^{M},\theta+\mu)$. By direct verification, one establishes
that the function
\begin{equation}
S_{E}\left(  (\mathcal{A},\partial_{\theta}\mathcal{A})(\theta),\theta\right)
\equiv\frac{\partial S_{\mathrm{L}}(\theta)}{\partial(\partial_{\theta}%
^{r}\mathcal{A}^{I}(\theta))}\partial_{\theta}^{r}\mathcal{A}^{I}%
(\theta)-S_{\mathrm{L}}(\theta) \label{13}%
\end{equation}
is an LS integral of motion, i.e., a conserved quantity under the $\theta
$-evolution, in case there holds the equation%
\begin{equation}
\left.  \frac{\partial}{\partial\theta}S_{\mathrm{L}}(\theta)+2(\partial
_{\theta}U)(\theta)S_{\mathrm{L}}(\theta)\right|  _{\mathcal{L}_{I}%
^{l}S_{\mathrm{L}}=0}=0. \label{14}%
\end{equation}
In contrast to its analogue in a $t$-local field theory, the energy $E(t)$,
the function $S_{E}(\theta)$ is an LS integral also in the case of an explicit
dependence on $\theta$. This fact takes place in case $S_{\mathrm{L}}(\theta)$
admits the structure%
\begin{equation}
S_{\mathrm{L}}\left(  (\mathcal{A},\partial_{\theta}\mathcal{A})(\theta
),\theta\right)  =S_{\mathrm{L}}^{0}\left(  \mathcal{A},\partial_{\theta
}\mathcal{A}\right)  (\theta)-2\theta(\partial_{\theta}U)(\theta
)S_{\mathrm{L}}^{0}(\theta),\;\vec{\varepsilon}(S_{\mathrm{L}}^{0})=\vec{0}.
\label{15}%
\end{equation}

\section{Odd-time Hamiltonian Formulation}

Independently, an LSM can be formulated in terms of a \emph{Hamiltonian
action}, being a $C^{\infty}(T_{\mathrm{odd}}^{\ast}\mathcal{M}_{\mathrm{CL}%
})$-function, $S_{\mathrm{H}}$: $T_{\mathrm{odd}}^{\ast}\mathcal{M}%
_{\mathrm{CL}}\times\{\theta\}\rightarrow\Lambda_{1}(\theta;\mathbb{R})$,
depending on superantifields $\mathcal{A}_{I}^{\ast}(\theta)=(A_{I}^{\ast
}-\theta J_{I})$, included in the local coordinates of $T_{\mathrm{odd}}%
^{\ast}\mathcal{M}_{\mathrm{CL}}$: $\Gamma_{\mathrm{CL}}^{P}(\theta
)=(\mathcal{A}^{I},\mathcal{A}_{I}^{\ast})(\theta)$, $\vec{\varepsilon
}(\mathcal{A}_{I}^{\ast})=\vec{\varepsilon}(\mathcal{A}^{I})+(1,0,1)$. The
equivalence of the Lagrangian and Hamiltonian formulations is implied by the
nondegeneracy of the supermatrix $\left\|  (S_{\mathrm{L}}^{\prime\prime
})_{IJ}(\theta)\right\|  $ given by (\ref{5}), in the framework of a Legendre
transformation of $S_{\mathrm{L}}(\theta)$ with respect to $\partial_{\theta
}^{r}\mathcal{A}^{I}(\theta)$,%
\begin{equation}
S_{\mathrm{H}}(\Gamma_{\mathrm{CL}}(\theta),\theta)=\mathcal{A}_{I}^{\ast
}(\theta)\partial_{\theta}^{r}\mathcal{A}^{I}(\theta)-S_{\mathrm{L}}%
(\theta),\;\mathcal{A}_{I}^{\ast}(\theta)=\frac{\partial S_{\mathrm{L}}%
(\theta)}{\partial(\partial_{\theta}^{r}\mathcal{A}^{I}(\theta))}, \label{16}%
\end{equation}
where $S_{\mathrm{H}}(\Gamma_{\mathrm{CL}}(\theta),\theta)$ coincides with
$S_{E}(\theta)$ in terms of the $T_{\mathrm{odd}}^{\ast}\mathcal{M}%
_{\mathrm{CL}}$-coordinates.

The dynamics of an LSM is given by a \emph{generalized Hamiltonian system} of
$3N$ first-order equations in $\theta$, equivalent to the LS equations in
(\ref{5}), and expressed through a $\theta$-local antibracket $(\,\cdot
\,,\,\cdot\,)_{\theta}$, namely,%
\begin{align}
&  \,\partial_{\theta}^{r}\Gamma_{\mathrm{CL}}^{P}(\theta)=\left(
\Gamma_{\mathrm{CL}}^{P}(\theta),S_{\mathrm{H}}(\theta)\right)  _{\theta
},\ \Theta_{I}^{\mathrm{H}}(\Gamma_{\mathrm{CL}}(\theta),\theta)=\Theta
_{I}(\mathcal{A}(\theta),\partial_{\theta}\mathcal{A}(\Gamma_{\mathrm{CL}%
}(\theta),\theta),\theta)=0,\nonumber\\
&  \,\left(  \mathcal{F}_{1}(\theta),\mathcal{F}_{2}(\theta)\right)  _{\theta
}\equiv\frac{\partial\mathcal{F}_{1}(\theta)}{\partial\mathcal{A}^{I}(\theta
)}\frac{\partial\mathcal{F}_{2}(\theta)}{\partial\mathcal{A}_{I}^{\ast}%
(\theta)}-\frac{\partial_{r}\mathcal{F}_{1}(\theta)}{\partial\mathcal{A}%
_{I}^{\ast}(\theta)}\frac{\partial_{l}\mathcal{F}_{2}(\theta)}{\partial
\mathcal{A}^{I}(\theta)},\ \mathcal{F}_{\mathrm{t}}(\theta)\in C^{\infty}(\Pi
T^{\ast}\mathcal{M}_{\mathrm{CL}}),\mathrm{t}=1,2 \label{17}%
\end{align}
with \emph{Hamiltonian constraints} $\Theta_{I}^{\mathrm{H}}(\Gamma
_{\mathrm{CL}}(\theta),\theta)$. The latter coincide with half of the
equations of the HS proper, due to transformations (\ref{16}) and their
consequences:%
\begin{equation}
\Theta_{I}^{\mathrm{H}}(\Gamma_{\mathrm{CL}}(\theta),\theta)=-\partial
_{\theta}^{r}\mathcal{A}_{I}^{\ast}(\theta)-S_{\mathrm{H}},_{I}(\theta
)(-1)^{\varepsilon_{I}}. \label{18}%
\end{equation}
Formula (\ref{18}) establishes the equivalence of an HS with a generalized HS,
and hence with an LS in the corresponding [formal, in view of the degeneracy
conditions (\ref{6})] setting ($\theta=0$, $k=\mathrm{CL}$) of the Cauchy
problem for integral curves $\hat{\mathcal{A}}^{I}(\theta)$, $\hat{\Gamma}%
{}_{k}^{P}(\theta)$,%
\begin{equation}
\left(  \hat{\mathcal{A}}^{I},\partial_{\theta}^{r}\hat{\mathcal{A}}%
^{I}\right)  (0)=\left(  \overline{\mathcal{A}}{}^{I},\overline{\partial
_{\theta}^{r}\mathcal{A}}{}^{I}\right)  ,\;\hat{\Gamma}_{k}^{P}(0)=\left(
\overline{\mathcal{A}}{}^{I},\overline{\mathcal{A}}{}_{I}^{\ast}\right)
:\;\overline{\mathcal{A}}{}_{I}^{\ast}=P_{0}\left[  \frac{\partial
S_{\mathrm{L}}(\theta)}{\partial(\partial_{\theta}^{r}\mathcal{A}^{I}%
(\theta))}\right]  \left(  \overline{\mathcal{A}}{}^{I},\overline
{\partial_{\theta}^{r}\mathcal{A}}{}^{I}\right)  , \label{19}%
\end{equation}
where we have ignored the continuous part of $I$. The equivalence between an
HS and a generalized HS is valid due to the coincidence (mutual inclusion) of
the corresponding sets of solutions. Indeed, the solutions of a generalized HS
are included into those of an HS by construction, while the reverse is valid
due to (\ref{18}).

The HS is defined through a variational problem for a functional identical to
$Z[\mathcal{A}]$,%
\begin{align}
&  Z_{\mathrm{H}}[{\Gamma}_{k}]=\int d\theta\left[  \frac{1}{2}{\Gamma}%
_{k}^{P}(\theta)\omega_{PQ}^{k}(\theta)\partial_{\theta}^{r}{\Gamma}_{k}%
^{Q}(\theta)-S_{\mathrm{H}}(\Gamma_{k}(\theta),\theta)\right]  ,\nonumber\\
&  \omega_{k}^{PQ}(\theta)\equiv\left(  \Gamma_{k}^{P}(\theta),\Gamma_{k}%
^{Q}(\theta)\right)  _{\theta},\;\omega_{k}^{PD}(\theta)\omega_{DQ}^{k}%
(\theta)=\delta^{P}{}_{Q}. \label{20}%
\end{align}
Definitions (\ref{9})--(\ref{11}) remain the same for special gauge theories,
while definitions (\ref{7}), (\ref{8}), in the case of general gauge theories
of $L_{g}$-stage reducibility, are transformed by the rule%
\begin{equation}
{\hat{\mathcal{Z}}}_{\mathrm{H}}\,_{\mathcal{A}_{s}}^{\mathcal{A}_{s-1}%
}\left(  \Gamma_{k}(\theta_{s-1}),\theta_{s-1};\theta_{s}\right)
={\hat{\mathcal{Z}}}_{\mathcal{A}_{s}}^{\mathcal{A}_{s-1}}\left(
\mathcal{A}(\theta_{s-1}),\partial_{\theta_{s-1}}\mathcal{A}(\Gamma_{k}%
(\theta_{s-1}),\theta_{s-1}),\theta_{s-1};\theta_{s}\right)
\,,\;s=0,...,L_{g}\,. \label{21}%
\end{equation}
Eqs. (\ref{14}), transformations (\ref{16}) and their consequence
$\frac{\partial}{\partial\theta}(S_{\mathrm{L}}+S_{\mathrm{H}})(\theta)=0$
imply the invariance of $S_{\mathrm{H}}(\theta)$ under $\theta$-shifts along
arbitrary solutions $\hat{\Gamma}{}_{k}^{P}(\theta)$, or, equivalently, along
an $(\varepsilon_{P},\varepsilon)$-odd vector field $\mathbf{Q}(\theta
)=\mathrm{ad}{}S_{\mathrm{H}}(\theta)\equiv(S_{\mathrm{H}}(\theta
),\,\cdot\,)_{\theta}$. Hence,%
\begin{equation}
\left.  \delta_{\mu}S_{\mathrm{H}}(\theta)\right|  _{\hat{\Gamma}_{k}(\theta
)}=\mu\left[  \frac{\partial}{\partial\theta}S_{\mathrm{H}}(\theta)-\left(
S_{\mathrm{H}}(\theta),S_{\mathrm{H}}(\theta)\right)  _{\theta}\right]
=0,\;\delta_{\mu}S_{\mathrm{H}}(\theta)=\mu\partial_{\theta}S_{\mathrm{H}%
}(\theta) \label{22}%
\end{equation}
holds true, provided that $S_{\mathrm{H}}(\theta)$ can be presented, according
to (\ref{14}), in the form%
\begin{equation}
S_{\mathrm{H}}\left(  \Gamma_{k}(\theta),\theta\right)  =S_{\mathrm{H}}%
^{0}\left(  \Gamma_{k}(\theta)\right)  +\theta\left(  S_{\mathrm{H}}%
^{0}\left(  \Gamma_{k}(\theta)\right)  ,S_{\mathrm{H}}^{0}\left(  \Gamma
_{k}(\theta)\right)  \right)  _{\theta}, \label{23}%
\end{equation}
where $(\partial_{\theta}U)(\theta)S_{\mathrm{L}}(\theta)=1/2\left(
S_{\mathrm{H}}(\theta),S_{\mathrm{H}}(\theta)\right)  _{\theta}$, and
$S_{\mathrm{H}}^{0}\left(  \Gamma_{k}(\theta)\right)  $ is the Legendre
transform of $S_{\mathrm{L}}^{0}(\theta)$, defined by (\ref{15}).

If $S_{\mathrm{H}}(\theta)$ or $S_{\mathrm{L}}(\theta)$ does not depend on
$\theta$ explicitly, then eq. (\ref{22}) or (\ref{14}) implies the fulfilment
of the equation $\left(  {S_{\mathrm{H}}(\theta),S_{\mathrm{H}}(\theta
)}\right)  {_{\theta}}=0$, or $\left.  (\partial_{\theta}U)(\theta
)S_{\mathrm{L}}(\theta)\right|  _{\hat{\mathcal{A}}(\theta)}=0$, which has no
counterpart in a $t$-local field theory, and imposes the known condition
\cite{BV} that $S_{\mathrm{H}}(\theta)$ or $S_{\mathrm{L}}(\theta)$ be proper,
although for an LSM at the classical level. In this case, a $\theta
$-superfield integrability\footnote{The notion of $\theta$-superfield
integrability is introduced by analogy with the treatment of Ref.
\cite{BatalinDamgaard}.} of the HS in (\ref{17}) is guaranteed by the standard
properties of the antibracket, including the Jacobi identity:%
\begin{equation}
(\partial_{\theta}^{r})^{2}\Gamma_{k}^{P}(\theta)={\frac{1}{2}}\left(
{\Gamma_{k}^{P}(\theta),}\left(  {S_{\mathrm{H}}(\Gamma_{k}(\theta
)),S_{\mathrm{H}}(\Gamma_{k}(\theta))}\right)  _{\theta}\right)  _{\theta}=0.
\label{24}%
\end{equation}
This fact ensures the validity on $C^{\infty}(T_{\mathrm{odd}}^{\ast
}\mathcal{M}_{\mathrm{CL}}\times\{\theta\})$ of the $\theta$-translation
formula%
\begin{equation}
\left.  \delta_{\mu}\mathcal{F}(\theta)\right|  _{\hat{\Gamma}_{k}(\theta
)}=\mu\left[  \frac{\partial}{\partial\theta}-\mathrm{ad}{}S_{\mathrm{H}%
}(\theta)\right]  \mathcal{F}(\theta)\equiv\mu\check{s}^{l}(\theta
)\mathcal{F}(\theta), \label{25}%
\end{equation}
as well as the nilpotency of a BRST-like generator of $\theta$-shifts along
$\mathbf{Q}(\theta)$, $\check{s}^{l}(\theta)$.

Depending on the realization of additional properties of a gauge theory (see
Section 4), we shall henceforth assume the validity of the equation%
\begin{equation}
\Delta^{k}(\theta)S_{\mathrm{H}}(\theta)=0,\;\Delta^{k}(\theta)\equiv
\frac{1}{2}(-1)^{\varepsilon(\Gamma^{Q})}\omega_{QP}^{k}(\theta)\left(
\Gamma_{k}^{P}(\theta),\left(  \Gamma_{k}^{Q}(\theta),\,\cdot\,\right)
_{\theta}\right)  _{\theta}. \label{26}%
\end{equation}
Eq. (\ref{26}) is equivalent to a vanishing divergence of the vector field
$\mathbf{Q}(\theta)$, namely,%
\begin{equation}
\mathrm{div}\,(\left.  \partial_{\theta}^{r}\Gamma_{k}(\theta)\right|
_{\hat{\Gamma}_{k}(\theta)})=\frac{\partial_{r}}{\partial\Gamma_{k}^{P}%
(\theta)}\left(  \left.  \partial_{\theta}^{r}\Gamma_{k}^{P}(\theta)\right|
_{\hat{\Gamma}_{k}(\theta)}\right)  =2\Delta^{k}(\theta)S_{\mathrm{H}}%
(\theta)=0. \label{27}%
\end{equation}
This condition is trivial for the symplectic counterpart of formula
(\ref{27}). The validity of the \emph{Hamiltonian master equation} $\left(
S_{\mathrm{H}}(\theta),S_{\mathrm{H}}(\theta)\right)  _{\theta}=0$ in case
$\frac{\partial}{\partial\theta}S_{\mathrm{H}}(\theta)=0$ justifies the
interpretation of the equivalent equation in (\ref{14}), for $\frac{\partial
}{\partial\theta}S_{\mathrm{L}}(\theta)=0$, $\left.  (\partial_{\theta
}U)(\theta)S_{\mathrm{L}}(\theta)\right|  _{\mathcal{L}_{I}^{l}S_{\mathrm{L}%
}=0}=0$, as a \emph{Lagrangian master equation}.

\section{Local Superfield Quantization}

In order to set up the rules of local superfield quantization for a gauge
model, we should first extract such a model from a general LSM. Then we should
consider a procedure of constructing a quantum action for the restricted LSM,
and, finally, investigate the possibility (inherent in the $\theta$-local
approach) of a \emph{dual description} of the LSM in terms of the quantities
of the BFV formalism.

\subsection{Superfield Quantum Action in Initial Coordinates}

In this subsection, we transform the reducibility relations of a
\emph{restricted\ }special LSM into a sequence of new gauge transformations
for the ghost superfields of the minimal sector. Together with the gauge
transformations of the classical superfields $\mathcal{A}^{i}(\theta)$,
extracted from $\mathcal{A}^{I}(\theta)$, the new gauge transformations are
translated into a Hamiltonian system related to the initial restricted HS. A
requirement of superfield integrability for the resulting HS produces a
deformation of the $\theta$-local Hamiltonian in powers of the ghosts and
superantifields of the minimal sector, and leads to a quantum action, and,
independently, to a gauge-fixing action (Subsection 4.3), subject to different
$\theta$-local master equations.

Given the standard distribution of ghost number \cite{BV} for $\Gamma
_{\mathrm{CL}}^{P}(\theta)$, $\mathrm{gh}(\mathcal{A}_{I}^{\ast}%
)=-1-\mathrm{gh}(\mathcal{A}^{I})=-1$, the choice $\mathrm{gh}(\theta
,\partial_{\theta})=(-1,1)$ implying the absence of ghosts among
$\mathcal{A}^{I}$, and, in particular, the relations $(\varepsilon_{P})_{I}%
=0$, the quantization rules consists, firstly, in restricting an LSM (in both
Lagrangian and Hamiltonian formulations) by the equations%
\begin{equation}
\left(  \mathrm{gh},\frac{\partial}{\partial\theta}\right)  S_{\mathrm{H}%
(\mathrm{L})}(\theta)=(0,0). \label{28}%
\end{equation}
Given the existence of a potential term in $S_{\mathrm{H}(\mathrm{L})}%
(\theta)$, $S(\mathcal{A}(\theta),0)=\mathcal{S}(\mathcal{A}(\theta))$, and
the absence in $S_{\mathrm{H}(\mathrm{L})}(\theta)$ of a dimensional constants
with a nonzero ghost number, solutions of eqs. (\ref{28}) select from an LSM a
usual gauge model with a classical action $S_{0}(A)$ in which the fields
$A^{i}$ are extended to $\mathcal{A}^{i}(\theta)$. Then the generalized HS in
(\ref{17}) is transformed into a $\theta$-integrable system defined in $\Pi
T^{\ast}\mathcal{M}_{\mathrm{cl}}=\{\Gamma_{\mathrm{cl}}^{p}(\theta
)\}=\{(\mathcal{A}^{i},\mathcal{A}_{i}^{\ast})(\theta)\}$, with $\Theta
_{i}^{\mathrm{H}}(\mathcal{A}(\theta))=\Theta_{i}(\mathcal{A}(\theta))$,
\begin{equation}
\partial_{\theta}^{r}\Gamma_{\mathrm{cl}}^{p}(\theta)=\left(  \Gamma
_{\mathrm{cl}}^{p}(\theta),S_{0}(\mathcal{A}(\theta))\right)  _{\theta
},\;\Theta_{i}^{\mathrm{H}}(\mathcal{A}(\theta))=-(-1)^{\varepsilon_{i}}%
S_{0},_{i}(\mathcal{A}(\theta)). \label{29}%
\end{equation}
The restricted special gauge transformations $\delta\mathcal{A}^{i}%
(\theta)=\mathcal{R}_{0\alpha_{0}}^{i}\left(  \mathcal{A}(\theta)\right)
{\xi}_{0}^{\alpha_{0}}(\theta)$, $\vec{\varepsilon}({\xi}_{0}^{\alpha_{0}%
}(\theta))=\vec{\varepsilon}_{\alpha_{0}}$, with the condition $(\varepsilon
_{P})_{\alpha_{0}}=0$, are embedded by the substitution $\xi_{0}^{\alpha_{0}%
}(\theta)=d\tilde{\xi}_{0}^{\alpha_{0}}(\theta)=\mathcal{C}^{\alpha_{0}%
}(\theta)d\theta$, $\alpha_{0}=1,...$, $m_{0}=m_{0-}+m_{0+}$, into a
Hamiltonian system with $2n$ equations for unknown $\Gamma_{\mathrm{cl}}%
^{p}(\theta)$, with the Hamiltonian $S_{1}^{0}(\Gamma_{\mathrm{cl}}%
,C_{0})(\theta)=(\mathcal{A}_{i}^{\ast}\mathcal{R}_{0}{}_{\alpha_{0}}%
^{i}(\mathcal{A})\mathcal{C}^{\alpha_{0}})(\theta)$. A union of this system
with the HS in (\ref{29}), extended to $2(n+m_{0})$ equations, has the form%
\begin{equation}
\partial_{\theta}^{r}\Gamma_{\lbrack0]}^{p_{[0]}}(\theta)=\left(
\Gamma_{\lbrack0]}^{p_{[0]}}(\theta),S_{[1]}^{0}(\theta)\right)  _{\theta
},\;S_{[1]}^{0}(\theta)=(S_{0}+S_{1}^{0})(\theta),\;\Gamma_{\lbrack
0]}^{p_{[0]}}\equiv(\Gamma_{\mathrm{cl}}^{p},\Gamma_{0}^{p_{0}}),\;\Gamma
_{0}^{p_{0}}\equiv(\mathcal{C}^{\alpha_{0}},\mathcal{C}_{\alpha_{0}}^{\ast}).
\label{30}%
\end{equation}
By virtue of (\ref{11}), the function $S_{1}^{0}(\theta)$ is invariant, modulo
$S_{0},_{i}(\theta)$, under special gauge transformations of ghost superfields
$\mathcal{C}^{\alpha_{0}}(\theta)$, with arbitrary functions $\xi_{1}%
^{\alpha_{1}}(\theta)$, $(\varepsilon_{P})_{\alpha_{1}}=0$, on the superspace
$\mathcal{M}$:%
\begin{equation}
\delta\mathcal{C}^{\alpha_{0}}(\theta)=\mathcal{Z}_{\alpha_{1}}^{\alpha_{0}%
}(\mathcal{A}(\theta))\xi_{1}^{\alpha_{1}}(\theta),\;(\vec{\varepsilon
},\mathrm{gh})\xi_{1}^{\alpha_{1}}(\theta)=\left(  \vec{\varepsilon}%
_{\alpha_{1}}+(1,0,1),1\right)  . \label{31}%
\end{equation}
Making the substitution $\xi_{1}^{\alpha_{1}}(\theta)=d\tilde{\xi}_{1}%
^{\alpha_{1}}(\theta)=\mathcal{C}^{\alpha_{1}}(\theta)d\theta$, $\alpha
_{1}=1,...,m_{1}$, and an enlargement of $m_{0}$ first-order equations in
$\theta$, with respect to the unknowns $\mathcal{C}^{\alpha_{0}}(\theta)$ in
transformations (\ref{31}), to an HS of $2m_{0}$ equations with the
Hamiltonian $S_{1}^{1}(\mathcal{A},\mathcal{C}_{0}^{\ast},\mathcal{C}%
_{1})(\theta)=(\mathcal{C}_{\alpha_{0}}^{\ast}\mathcal{Z}_{\alpha_{1}}%
^{\alpha_{0}}(\mathcal{A})\mathcal{C}^{\alpha_{1}})(\theta)$, we obtain a
system of the form (\ref{30}), written for $\partial_{\theta}^{r}\Gamma
_{0}^{p_{0}}(\theta)$. The enlargement of the union of the latter HS with eqs.
(\ref{30}) is formally identical to the system (\ref{30}) under the
replacement%
\[
(\Gamma_{\lbrack0]}^{p_{[0]}},S_{[1]}^{0})\rightarrow(\Gamma_{\lbrack
1]}^{p_{[1]}},S_{[1]}^{1}):\;\left\{  \Gamma_{\lbrack1]}^{p_{[1]}}%
=(\Gamma_{\lbrack0]}^{p_{[0]}},\Gamma_{1}^{p_{1}}),\;\Gamma_{1}^{p_{1}%
}=(\mathcal{C}^{\alpha_{1}},\mathcal{C}_{\alpha_{1}}^{\ast}),\;S_{[1]}%
^{1}=S_{[1]}^{0}+S_{1}^{1}\right\}  .
\]

The iteration sequence related to a reformulation of the special gauge
transformations of ghosts $\mathcal{C}^{\alpha_{0}}$,$\ldots$, $\mathcal{C}%
^{\alpha_{s-2}}$, obtained from (possibly) enhanced\footnote{From
$\mathrm{gh}(\mathcal{A}^{I})=0$ in eqs. (\ref{28}), with $(\varepsilon
_{P})_{\mathcal{A}_{s}}=(\varepsilon_{P})_{I}=0$, $s=0,...,L_{g}$, it follows
that the values of $\overline{m}$, $\overline{m}_{s}$ may be both larger and
smaller than the corresponding values $\overline{M}$, $\overline{M}_{s}$, in
contrast to the values of $\overline{n}$, $\overline{N}$. Indeed, for a
restricted LSM, the presence of additional gauge symmetries is possible;
therefore, we suppose that (possibly) enhanced sets of restricted functions
$\mathcal{R}_{0}{}_{\alpha_{0}}^{i}(\theta)$, $\mathcal{Z}_{\alpha_{s}%
}^{\alpha_{s-1}}(\theta)$ exhaust, correspondingly, on the surface $S_{0}%
,_{i}(\theta)=0$, the zero-modes of both the Hessian $S_{0},_{ij}(\theta)$ and
$\mathcal{Z}_{\alpha_{s-1}}^{\alpha_{s-2}}(\theta)$. As a consequence, this
implies that the final stage of reducibility for a restricted model $L$ is
different from $L_{g}$.} relations (\ref{11}), leads, for an $L$%
-stage-reducible restricted LSM at the $s$-th step with $0<s\leq L$ and
$\Gamma_{\mathrm{cl}}^{p}\equiv\Gamma_{-1}^{p_{-1}}$, to invariance
transformations for $S_{1}^{s-1}(\theta)$, modulo $S_{0},_{i}(\theta)$,
namely,%
\begin{align}
&  \delta\mathcal{C}^{\alpha_{s-1}}(\theta)=\mathcal{Z}_{\alpha_{s}}%
^{\alpha_{s-1}}(\mathcal{A}(\theta))\xi_{s}^{\alpha_{s}}(\theta),\;(\vec
{\varepsilon},\mathrm{gh})\xi_{s}^{\alpha_{s}}(\theta)=(\vec{\varepsilon
}_{\alpha_{s}}+s(1,0,1),s),\;(\varepsilon_{P})_{\alpha_{s}}=0,\nonumber\\
&  S_{1}^{s-1}(\theta)=(\mathcal{C}_{\alpha_{s-2}}^{\ast}\mathcal{Z}%
_{\alpha_{s-1}}^{\alpha_{s-2}}(\mathcal{A})\mathcal{C}^{\alpha_{s-1}}%
)(\theta),\;\left(  \mathrm{gh},\frac{\partial}{\partial\theta}\right)
S_{1}^{s-1}(\theta)=(0,0). \label{32}%
\end{align}
The substitution $\xi_{s}^{\alpha_{s}}(\theta)=d\tilde{\xi}_{s}^{\alpha_{s}%
}(\theta)=\mathcal{C}^{\alpha_{s}}(\theta)d\theta$, $\alpha_{s}=1,...$,
$m_{s}=m_{s-}+m_{s+}$, transforms special gauge transformations (\ref{32})
into $m_{s-1}$ equations for unknown $\mathcal{C}^{\alpha_{s-1}}(\theta)$,
extended by the introduction of superantifields $\mathcal{C}_{\alpha_{s-1}%
}^{\ast}(\theta)$\ to an HS:%
\begin{equation}
\partial_{\theta}^{r}\Gamma_{s-1}^{p_{s-1}}(\theta)=\left(  \Gamma
_{s-1}^{p_{s-1}}(\theta),S_{1}^{s}(\theta)\right)  _{\theta},\;S_{1}%
^{s}(\theta)=(\mathcal{C}_{\alpha_{s-1}}^{\ast}\mathcal{Z}_{\alpha_{s}%
}^{\alpha_{s-1}}(\mathcal{A})\mathcal{C}^{\alpha_{s}})(\theta),\;\Gamma
_{s-1}^{p_{s-1}}=(\mathcal{C}^{\alpha_{s-1}},\mathcal{C}_{\alpha_{s-1}}^{\ast
}). \label{33}%
\end{equation}
Having combined the system (\ref{33}) with an HS of the same form, although
with $\partial_{\theta}^{r}\Gamma_{\lbrack s-1]}^{p_{[s-1]}}(\theta)$ and the
Hamiltonian $S_{[1]}^{s-1}(\theta)=(S_{0}+\sum_{r=0}^{s-1}S_{1}^{r})(\theta)$,
and having expressed the result for $2\left(  n+\sum_{r=0}^{s}m_{r}\right)  $
equations with $S_{[1]}^{s}(\theta)=(S_{[1]}^{s-1}+S_{1}^{s})(\theta)$, we
obtain, by induction, the following HS:%
\begin{equation}
\partial_{\theta}^{r}\Gamma_{\lbrack L]}^{p_{[L]}}(\theta)=\left(
\Gamma_{\lbrack L]}^{p_{[L]}}(\theta),S_{[1]}^{L}(\theta)\right)  _{\theta
}\,,\;S_{[1]}^{L}(\theta)=S_{0}(\mathcal{A}(\theta))+\sum\limits_{s=0}%
^{L}(\mathcal{C}_{\alpha_{s-1}}^{\ast}\mathcal{Z}_{\alpha_{s}}^{\alpha_{s-1}%
}(\mathcal{A})\mathcal{C}^{\alpha_{s}})(\theta)\,. \label{34}%
\end{equation}
The function $S_{[1]}^{L}(\theta)$, subject to the condition of a proper
$\theta$-local solution of the classical master equation \cite{BV}, with the
antibracket extended in $\Pi T^{\ast}\mathcal{M}_{k}$=$\{\Gamma_{\lbrack
L]}^{p_{[L]}}(\theta)\equiv\Gamma_{k}^{p_{k}}(\theta)$=$(\Phi^{A_{k}}%
,\Phi_{A_{k}}^{\ast})(\theta)$, $A_{k}=1,\ldots,n+\sum_{r=0}^{L}m_{r}$%
,$\;k$=$\mathrm{min}\}$, is a solution of the master equation with accuracy up
to $O(C^{\alpha_{s}})$, modulo $S_{0},_{i}(\theta)$. The integrability of the
HS in (\ref{34}) is guaranteed by a double deformation of $S_{[1]}^{L}%
(\theta)$: first in powers of $\Phi_{A_{k}}^{\ast}(\theta)$ and then in powers
of $\mathcal{C}^{\alpha_{s}}(\theta)$, in the framework of the existence
theorem \cite{BV1} for the classical master equation in the minimal sector:%
\begin{equation}
\left(  S_{\mathrm{H};k}(\Gamma_{k}(\theta)),S_{\mathrm{H};k}(\Gamma
_{k}(\theta))\right)  _{\theta}=0,\;\left(  \vec{\varepsilon},\mathrm{gh}%
,\frac{\partial}{\partial\theta}\right)  S_{\mathrm{H};k}(\Gamma_{k}%
(\theta))=\left(  \vec{0},0,0\right)  ,\;k=\mathrm{min}. \label{35}%
\end{equation}
The proposed superfield algorithm for constructing the function $S_{\mathrm{H}%
;\mathrm{min}}(\theta)$ may be considered as a superfield version of the
Koszul--Tate complex resolution \cite{KoszulTate}.

Let us consider an extension of $S_{\mathrm{H};\mathrm{min}}(\theta)$ to
$S_{\mathrm{H};k}(\theta)=S_{\mathrm{H};\mathrm{min}}(\theta)+\sum_{s=0}%
^{L}\sum_{s^{\prime}=0}^{s}(\mathcal{C}_{s^{\prime}{}\alpha_{s}}^{\ast
}\mathcal{B}_{s^{\prime}}^{\alpha_{s}})(\theta)$, being a proper solution
\cite{BV} in $\Pi T^{\ast}\mathcal{M}_{k}=\{\Gamma_{k}^{p_{k}}(\theta)\}$,%
\begin{align}
&  \Gamma_{k}^{p_{k}}(\theta)=(\Gamma_{\mathrm{min}}^{p_{\mathrm{min}}%
},\mathcal{C}_{s^{\prime}}^{\alpha_{s}},\mathcal{B}_{s^{\prime}}^{\alpha_{s}%
},\mathcal{C}_{s^{\prime}{}\alpha_{s}}^{\ast},\mathcal{B}_{s^{\prime}{}%
\alpha_{s}}^{\ast})(\theta),\;s^{\prime}=0,...,s,\;s=0,...,L,\nonumber\\
&  (\vec{\varepsilon},\mathrm{gh})\mathcal{C}_{s^{\prime}}^{\alpha_{s}}%
(\theta)=(\vec{\varepsilon}_{\alpha_{s}}+(s+1)(1,0,1),2s^{\prime}%
-s-1)=(\vec{\varepsilon},\mathrm{gh})\mathcal{B}_{s^{\prime}}^{\alpha_{s}%
}(\theta)+((1,0,1),-1)\nonumber
\end{align}
[henceforth we assume $k=\mathrm{ext}$ and take into account that
$(\vec{\varepsilon},\mathrm{gh})\Phi_{A_{k}}^{\ast}(\theta)=-((1,0,1),1)-(\vec
{\varepsilon},\mathrm{gh})\Phi^{A_{k}}(\theta)$], with the pyramids of ghosts
and Nakanishi--Lautrup superfields, and with a deformation in the Planck
constant $\hbar$. Then $S_{\mathrm{H};k}(\theta)$ determines the quantum
action $S_{\mathrm{H}}^{\Psi}(\Gamma(\theta),\hbar)$, e.g., in case of an
Abelian hypergauge defined as an anticanonical phase transformation:%
\begin{equation}
\Gamma_{k}^{p_{k}}(\theta)\rightarrow{\Gamma^{\prime}}_{k}^{p_{k}}%
(\theta)=\left(  \Phi^{A_{k}}(\theta),{\Phi}_{A_{k}}^{\ast}(\theta
)-\frac{\partial\Psi(\Phi(\theta))}{\partial\Phi^{A_{k}}(\theta)}\right)
:\;S_{\mathrm{H}}^{\Psi}(\Gamma(\theta),\hbar)=e^{\mathrm{ad}{}{\Psi}%
}S_{\mathrm{H};k}(\Gamma_{k}(\theta),\hbar). \label{36}%
\end{equation}
The functions $(S_{\mathrm{H}}^{\Psi},S_{\mathrm{H};k})(\theta,\hbar)$ obey
eqs. (\ref{26}), (\ref{35}) in case the $\hbar$-deformation of $S_{\mathrm{H}%
;\mathrm{min}}(\theta)$ is a solution of these equations. It is known that
this choice of equations ensures the integrability of a non-equivalent HS
constructed from $S_{\mathrm{H}}^{\Psi}$, $S_{\mathrm{H};k}$, as well as the
anticanonical [preserving the volume element $dV_{k}(\theta)=\prod_{p_{k}%
}d\Gamma_{k}^{p_{k}}(\theta)$] nature of this change of variables,
corresponding to a $\theta$-shift by a constant parameter $\mu$ along the
corresponding HS solutions. In its turn, the quantum master equation%
\begin{equation}
\Delta^{k}(\theta)\exp\left[  \frac{i}{\hbar}E(\theta,\hbar)\right]
=0,\;E\in\{S_{\mathrm{H}}^{\Psi},S_{\mathrm{H};k}\} \label{37}%
\end{equation}
determines a non-integrable HS, with the respective anticanonical change of
variables preserving $d\hat{V}_{k}(\theta)=\exp\left[  \left(  i/\hbar\right)
E(\theta,\hbar)\right]  dV_{k}(\theta)$. It is the latter nonintegrable HS
with the Hamiltonian $S_{\mathrm{H}}^{\Psi}(\theta,\hbar)$ that is crucial,
for $\theta=0$, in the BV formalism. This HS defines in $\Pi T^{\ast
}\mathcal{M}_{k}$ a $\theta$-local, but not nilpotent, generator of BRST
transformations, $\tilde{s}^{l(\Psi)}(\theta)$, which is associated with its
$\theta$-nonintegrable consequence:%
\begin{equation}
\partial_{\theta}^{r}\left(  {\Phi}^{A_{k}},{\Phi}_{A_{k}}^{\ast}\right)
(\theta)=\left(  \left(  \Phi^{A_{k}}(\theta),S_{\mathrm{H}}^{\Psi}%
(\theta,\hbar)\right)  _{\theta},0\right)  ,\;\tilde{s}^{l(\Psi)}%
(\theta)=\frac{\partial}{\partial\theta}+\frac{\partial_{r}S_{\mathrm{H}%
}^{\Psi}(\theta,\hbar)}{\partial\Phi_{A_{k}}^{\ast}(\theta)}\frac{\partial
_{l}}{\partial\Phi^{A_{k}}(\theta)}. \label{38}%
\end{equation}

\subsection{Duality between the BV and BFV Superfield Quantities}

An embedding of a restricted LSM gauge algebra, described by $S_{\mathrm{H}%
;\mathrm{min}}(\theta)$ and eq. (\ref{35}), into the gauge algebra of a
general gauge theory in the Lagrangian formalism, see eqs. (\ref{7}%
)--(\ref{12}), can be effectively realized by means of dual functional
counterparts, with the opposite ($\varepsilon_{P},\varepsilon$)-parity, of the
action and antibracket, by analogy with the approach of Refs.
\cite{AlexandrovKontsevichSchwarzZaboronsky,GrigorievDamgaard}. To this end,
let us consider the functional%
\[
Z_{k}[\Gamma_{k}]=-\partial_{\theta}S_{\mathrm{H};k}(\theta)\,,\;(\vec
{\varepsilon},\mathrm{gh})Z_{k}=((1,0,1),1)
\]
on the supermanifold $\Pi T(\Pi T^{\ast}\mathcal{M}_{k})=\{(\Gamma_{k}^{p_{k}%
},\partial_{\theta}\Gamma_{k}^{p_{k}})(\theta),k=\mathrm{min}\}$ with natural
($\varepsilon_{P},\varepsilon$)-even, symplectic, and ($\varepsilon
_{P},\varepsilon$)-odd Poisson structures. These structures define an
($\varepsilon_{P},\varepsilon$)-even functional $\{\,\cdot\,,\,\cdot\,\}$ with
canonical pairs $\{(\Phi_{k}^{A_{k}},\partial_{\theta}\Phi_{A_{k}}^{\ast})$,
$(\partial_{\theta}\Phi_{k}^{A_{k}},\Phi_{A_{k}}^{\ast})\}(\theta)$, and
($\varepsilon_{P},\varepsilon$)-odd $\theta$-local, $(\,\cdot\,,\,\cdot
\,)_{\theta}^{(\Gamma_{k},\partial_{\theta}\Gamma_{k})}$, Poisson brackets.
The latter act on the superalgebra $C^{\infty}(\Pi T(\Pi T^{\ast}%
\mathcal{M}_{k})\times\theta)$ and provide a lifting of the antibracket
$(\,\cdot\,,\,\cdot\,)_{\theta}$ defined on $\Pi T^{\ast}\mathcal{M}_{k}$. For
arbitrary functionals $F_{\mathrm{t}}[\Gamma_{k}]=\partial_{\theta}%
\mathcal{F}_{\mathrm{t}}\left(  (\Gamma_{k},\partial_{\theta}\Gamma
_{k})(\theta),\theta\right)  $, $\mathrm{t}=1,2$, one has the following
correspondence between the Poisson brackets of opposite Grassmann grading:%
\begin{align}
&  \hspace{-1em}\left\{  F_{1},F_{2}\right\}  =\int d\theta\left[
\frac{\delta F_{1}}{\delta\Phi^{A_{k}}(\theta)}\frac{\delta F_{2}}{\delta
\Phi_{A_{k}}^{\ast}(\theta)}-\frac{\delta_{r}F_{1}}{\delta\Phi_{A_{k}}^{\ast
}(\theta)}\frac{\delta_{l}F_{2}}{\delta\Phi^{A_{k}}(\theta)}\right]  =\int
d\theta(\mathcal{F}_{1}(\theta),\mathcal{F}_{2}(\theta))_{\theta}^{(\Gamma
_{k},\partial_{\theta}\Gamma_{k})},\nonumber\\
&  \hspace{-1em}(\mathcal{F}_{1}(\theta),\mathcal{F}_{2}(\theta))_{\theta
}^{(\Gamma_{k},\partial_{\theta}\Gamma_{k})}\equiv\left[  \left(
\mathcal{L}_{A_{k}}\mathcal{F}_{1}\right)  \mathcal{L}^{\ast A_{k}}%
\mathcal{F}_{2}-(\mathcal{L}_{r}^{\ast A_{k}}\mathcal{F}_{1})\mathcal{L}%
_{A_{k}}^{l}\mathcal{F}_{2}\right]  (\theta), \label{39}%
\end{align}
where the Euler--Lagrange superfield derivative, e.g., with respect to
$\Phi_{A_{k}}^{\ast}(\theta)$, for a fixed $\theta$, has the form
$\mathcal{L}^{\ast A_{k}}(\theta)=\partial/\partial\Phi_{A_{k}}^{\ast}%
(\theta)-(-1)^{\varepsilon_{A_{k}}+1}\partial_{\theta}\cdot\partial
/\partial\left(  \partial_{\theta}\Phi_{A_{k}}^{\ast}(\theta)\right)  $.

By construction, the functional $Z_{k}$ is nilpotent:%
\begin{equation}
\left\{  Z_{k},Z_{k}\right\}  =\int d\theta(S_{\mathrm{H};k}(\theta
),S_{\mathrm{H};k}(\theta))_{\theta}=0,\;k=\mathrm{min}, \label{40}%
\end{equation}
and, due to the absence of the additional time coordinate, is formally related
to the BRST charge of a dynamical system with first-class constraints
\cite{BFV}. Indeed, after identifying the fields $(\Gamma_{k},\partial
_{\theta}\Gamma_{k})(0)$ with the phase-space coordinates of the minimal
sector, canonical with respect to the $(\varepsilon_{P},\varepsilon)$-even
brackets in the framework of the BFV method \cite{BFV} for first-class
constrained systems of $(L+1)$-stage reducibility,%
\begin{align}
&  (q^{i},p_{i})=(\mathcal{A}^{i},\partial_{\theta}\mathcal{A}_{i}^{\ast
})(0),\;\left(  C^{A_{s}},\mathcal{P}_{A_{s}}\right)  =\left(  (\partial
_{\theta}^{r}\mathcal{C}^{\alpha_{s-1}},\mathcal{C}^{\alpha_{s}}%
),(\mathcal{C}_{\alpha_{s-1}}^{\ast},\partial_{\theta}\mathcal{C}_{\alpha_{s}%
}^{\ast})\right)  (0),\nonumber\\
&  A_{s}=(\alpha_{s-1},\alpha_{s}),\ s=0,...,L,\;\left(  C^{A_{L+1}%
},\mathcal{P}_{A_{L+1}}\right)  =\left(  \partial_{\theta}^{r}\mathcal{C}%
^{\alpha_{L}},\mathcal{C}_{\alpha_{L}}^{\ast}\right)  (0), \label{41}%
\end{align}
the functional $Z_{k}$ takes the form%
\begin{equation}
Z_{k}[\Gamma_{k}]=T_{A_{0}}(q,p)C^{A_{0}}+\sum\limits_{s=1}^{L+1}%
\mathcal{P}_{A_{s-1}}Z_{A_{s}}^{A_{s-1}}(q)C^{A_{s}}+O(C^{2}). \label{42}%
\end{equation}
With allowance for the gauge algebra structure functions of the original
$L$-stage-reducible restricted LSM described by the enhanced eqs. (\ref{11}),
the constraints $T_{A_{0}}(q,p)$ and the set of $(L+1)$-stage-reducible
eigenvectors $Z_{A_{s}}^{A_{s-1}}(q)$ are defined\ by the relations (the
symbol $T$ below stands for transposition)%
\begin{align}
&  \hspace{-2em}T_{A_{0}}(q,p)=\left(  S_{0},_{i}(q),-p_{i}\mathcal{R}_{0}%
{}_{\alpha_{0}}^{i}(q)\right)  ,\;Z_{A_{s}}^{A_{s-1}}(q)=\mathrm{diag}\left(
\mathcal{Z}_{\alpha_{s-1}}^{\alpha_{s-2}},\mathcal{Z}_{\alpha_{s}}%
^{\alpha_{s-1}}\right)  (q),\nonumber\\
&  \hspace{-2em}s=1,...,L,\;\left(  Z_{A_{L+1}}^{A_{L}}\right)  ^{T}%
(q)=\left(  \mathcal{Z}_{\alpha_{L}}^{\alpha_{L-1}},0\right)  ^{T}%
(q),\label{43}\\
&  \hspace{-2em}Z_{A_{s-1}}^{A_{s-2}}Z_{A_{s}}^{A_{s-1}}=T_{B_{0}}L_{A_{s}%
}^{A_{s-2}B_{0}}(q,p),\;s=1,...,\;L+1,\;Z_{A_{0}}^{A_{-1}}\equiv T_{A_{0}%
},\ L_{A_{s}}^{A_{s-2}\beta_{0}}=0,\nonumber\\
&  \hspace{-2em}L_{A_{s}}^{A_{s-2}j}=\mathrm{diag\,}\left(  \mathcal{L}%
_{\alpha_{s-1}}^{\alpha_{s-3}j},\;\mathcal{L}_{\alpha_{s}}^{\alpha_{s-2}%
j}\right)  ,\;\mathcal{L}_{\alpha_{0}}^{\alpha_{-2}j}=\mathcal{L}%
_{\alpha_{L+1}}^{\alpha_{L-1}j}=0,\;\mathcal{L}_{\alpha_{1}}^{\alpha_{-1}%
j}(q,p)=(-1)^{\varepsilon_{j}+1}p_{i}\mathcal{K}_{\alpha_{1}}^{ji}(q).
\label{44}%
\end{align}
Relations (\ref{39})--(\ref{44}) generalize, to the case of arbitrary
reducible theories, the results of Ref. \cite{GrigorievDamgaard} concerning a
dual description (for $\varepsilon_{i}=\varepsilon_{\alpha_{0}}=L=0$) of the
quantum action and classical master equation in terms of a nilpotent BRST charge.

By the rule (\ref{41}), the variables $(\mathcal{C}_{{s^{\prime}}\alpha_{s}%
}^{\ast},\mathcal{B}_{{s^{\prime}}\alpha_{s}}^{\ast},\mathcal{B}_{s^{\prime}%
}^{\alpha_{s}})(\theta)$ are identical to the respective ghost momenta
$\mathcal{P}_{{s^{\prime}}A_{s}}$, Lagrangian multipliers $\lambda
_{{s^{\prime}}A_{s}}$, and their conjugate momenta $\pi_{s^{\prime}}^{A_{s}}$
in \cite{BFV}. Then a comparison of the superfields $\mathcal{C}_{s^{\prime}%
}^{\alpha_{s}}(\theta)$, $s^{\prime}=0,...,s$, selected from the non-minimal
configuration space of an $L$-stage-reducible LSM, with the coordinates
$C_{s^{\prime}}^{A_{s}}$ selected from the non-minimal phase space of the
corresponding $(L+1)$-stage-reducible dynamical system \cite{BFV} demonstrates
the only possible embedding of $\Pi T(\Pi T^{\ast}\mathcal{M}_{\mathrm{ext}})$
into the phase space of the BFV method. Indeed, for the coordinates
$C_{0}^{A_{L+1}}$, $\mathrm{gh}(C_{0}^{A_{L+1}})=-L-2$, there exists no
pre-image among $(\mathcal{C}_{s^{\prime}}^{\alpha_{s}},\partial_{\theta
}\mathcal{C}_{s^{\prime}}^{\alpha_{s}})(0)$, because the ghost number spectrum
for the latter variables is bounded from below:%
\[
\min\mathrm{gh}(\mathcal{C}_{s^{\prime}}^{\alpha_{s}},\partial_{\theta
}\mathcal{C}_{s^{\prime}}^{\alpha_{s}})=\mathrm{gh}(\mathcal{C}_{0}%
^{\alpha_{L}})=-L-1.
\]
As a consequence, the nilpotent functional $Z_{k}[\Gamma_{k}]=-\partial
_{\theta}S_{\mathrm{H};k}(\theta)$, $k=\mathrm{ext}$, is embedded into the
total BRST charge constructed by the prescription of Ref. \cite{BFV}.

It should be noted that the systems constructed with respect to the
Hamiltonians $S_{\mathrm{H}}^{\Psi}(\Gamma(\theta),\hbar)$ and $S_{\mathrm{H}%
;k}(\theta)$, $k=\mathrm{min}$, $\mathrm{ext}$, are equivalently described by
dual fermion functionals $Z_{k}[\Gamma_{k}]$ and $Z^{\Psi}[\Gamma
]=-\partial_{\theta}S_{\mathrm{H}}^{\Psi}(\Gamma(\theta),\hbar)$, in terms of
even Poisson brackets, for instance,%
\begin{equation}
\partial_{\theta}^{r}\Gamma^{p}(\theta)=\left(  \Gamma^{p}(\theta
),S_{\mathrm{H}}^{\Psi}(\Gamma(\theta),\hbar)\right)  _{\theta}=-\left\{
\Gamma^{p}(\theta),Z^{\Psi}[\Gamma]\right\}  . \label{45}%
\end{equation}
Thereby, BRST transformations in the Lagrangian formalism with Abelian
hypergauges can be encoded by a formal BRST charge, $Z^{\Psi}[\Gamma]$,
related to $Z_{k}[\Gamma_{k}]$, $k=\mathrm{ext}$, by a phase canonical
transformation with the $(\varepsilon_{P},\varepsilon)$-even phase $F^{\Psi
}[\Phi]$=$\partial_{\theta}\Psi(\Phi(\theta))$,
\begin{equation}
Z^{\Psi}[\Gamma]=e^{\overline{\mathrm{ad}}\,F^{\Psi}}Z_{k}[\Gamma
_{k}]\,,\;\overline{\mathrm{ad}}\,F^{\Psi}\equiv\left\{  F^{\Psi}%
,\cdot\right\}  . \label{46}%
\end{equation}
On the assumption that an additional gauge invariance does not appear in
deriving the restricted LSM from the initial general gauge theory, i.e.,
$\overline{m}_{s}\leq\overline{M}_{s}$, and, therefore, $L\leq L_{g}$, cf.
footnote \thefootnote                          , the problem of including the
restricted LSM gauge algebra into the initial gauge algebra, defined by
(\ref{2}), (\ref{7}), (\ref{8}), is solved with the help of a nilpotent
functional defined on $\Pi T(\Pi T^{\ast}\mathcal{M}_{k})$=$\{(\Gamma
_{k}^{P_{k}},\partial_{\theta}\Gamma_{k}^{P_{k}})(\theta)$, $\Gamma_{k}%
^{P_{k}}(\theta)=\left(  \Gamma_{\mathrm{CL}}^{P_{\mathrm{CL}}},\mathcal{C}%
^{\mathcal{A}_{s}},\mathcal{C}_{\mathcal{A}_{s}}^{\ast}\right)  (\theta)$,
$s=0,1,...,L_{g}$, $k=\mathrm{MIN}\}$, namely,
\begin{align}
\hat{Z}_{k}[\Gamma_{k}]  &  =Z[\mathcal{A}]+\sum\limits_{s=0}^{L_{g}}\left[
\int d\theta_{s-1}d\theta_{s}\mathcal{C}_{\mathcal{A}_{s-1}}^{\ast}%
(\theta_{s-1})\hat{\mathcal{Z}}_{\mathcal{A}_{s}}^{\mathcal{A}_{s-1}}%
(\theta_{s-1};\theta_{s})\mathcal{C}^{\mathcal{A}_{s}}(\theta_{s}%
)(-1)^{\varepsilon_{\mathcal{A}_{s-1}}+s}+O(C^{\mathcal{A}_{s}})\right]
\nonumber\\
&  =\int d\theta S_{\mathrm{L};k}\left(  (\Gamma_{k},\partial_{\theta}%
\Gamma_{k})(\theta),\theta\right)  . \label{47}%
\end{align}

Given the superfields $\mathcal{C}^{\mathcal{A}_{s}}$ introduced as simple
ghosts $\mathcal{C}^{\alpha_{s}}$, although used for a description of a
general gauge algebra, a representation of solutions to the generating
equation $\left\{  \hat{Z}_{k},\hat{Z}_{k}\right\}  =0$ as expansions in
powers of $\mathcal{C}^{\mathcal{A}_{s}}$ can be controlled by an additional
\emph{generalized ghost number}, $\mathrm{gh}_{g}$, $\mathrm{gh}_{g}(\hat
{Z}_{k})=0$, coinciding with the standard ghost number only in the sector of
$(\Phi^{A_{\mathrm{MIN}}},\Phi_{A_{\mathrm{MIN}}}^{\ast})(0)$, for
$\mathrm{gh}(\mathcal{A}^{I},C^{\mathcal{A}_{s}})=(0,1+s)$, and having the
spectrum%
\[
\mathrm{gh}_{g}(\mathcal{A}^{I},C^{\mathcal{A}_{s}})=(0,1+s),\;\mathrm{gh}%
_{g}(\Phi_{A_{\mathrm{MIN}}}^{\ast})=-1-\mathrm{gh}_{g}(\Phi^{A_{\mathrm{MIN}%
}}),\;\mathrm{gh}_{g}(\theta,\partial_{\theta})=(0,0).
\]
Conditions (\ref{28}), applied to $S_{\mathrm{L};k}(\theta)$ in case
$(\varepsilon_{P})_{\mathcal{A}_{s}}=(\varepsilon_{P})_{I}=0$, $s=0,...,L_{g}%
$, extract from $\hat{Z}_{k}$ the functional ${Z}_{k}$ in (\ref{42}), so that
the ($\varepsilon_{P},\varepsilon$)-even $\theta$-density $S_{\mathrm{L}%
;k}(\theta)$ lifts the function $S_{\mathrm{H};k}(\theta)\in C^{\infty}(\Pi
T^{\ast}\mathcal{M}_{\mathrm{min}})$ to the superalgebra $C^{\infty}(\Pi T(\Pi
T^{\ast}\mathcal{M}_{\mathrm{MIN}})\times\theta)$. In general, $S_{\mathrm{L}%
;k}(\theta)$ does not obey the generalized master equation (\ref{35}) with the
antibracket (\ref{39}) acting on $C^{\infty}(\Pi T(\Pi T^{\ast}\mathcal{M}%
_{\mathrm{MIN}})\times\theta)$,
\begin{equation}
(S_{\mathrm{L};k}(\theta),S_{\mathrm{L};k}(\theta))_{\theta}^{(\Gamma
_{k},\partial_{\theta}\Gamma_{k})}=\tilde{f}\left(  (\Gamma_{k},\partial
_{\theta}\Gamma_{k})(\theta),\theta\right)  ,\;\tilde{f}(\theta)\in
\ker\{\partial_{\theta}\},\;k=\mathrm{MIN}. \label{48}%
\end{equation}

\subsection{Local Quantization}

Leaving aside the realization of a reducible LSM on $\Pi T^{\ast}%
\mathcal{M}_{\mathrm{ext}}$, we now suppose that the model is described by a
quantum action, $W(\theta,\hbar)=W(\theta)$, defined on an arbitrary
antisymplectic manifold $\mathcal{N}$ without connection, $\dim\mathcal{N}%
=\dim\Pi T^{\ast}\mathcal{M}_{\mathrm{ext}}=\overline{n}+(n_{-},n_{+}%
)+\sum_{r=0}^{L}(2r+3)(\overline{m}_{r}+(m_{r-},m_{r+}))$, with local
coordinates $\Gamma^{p}(\theta)$ and a density function $\rho(\Gamma(\theta
))$. A local antibracket, an invariant volume element, $d\mu(\Gamma(\theta))$,
and a nilpotent second-order operator, $\Delta^{\mathcal{N}}(\theta)$, are
defined in terms of an ($\varepsilon_{P},\varepsilon$)-odd Poisson bivector,
$\omega^{pq}(\Gamma(\theta))=\left(  \Gamma^{p}(\theta),\ \Gamma^{q}%
(\theta)\right)  _{\theta}^{\mathcal{N}}$, namely,%
\begin{equation}
d\mu(\Gamma(\theta))=\rho(\Gamma(\theta))d\Gamma(\theta),\;\Delta
^{\mathcal{N}}(\theta)=\frac{1}{2}(-1)^{\varepsilon(\Gamma^{q})}\rho
^{-1}\omega_{qp}(\theta)\left(  \Gamma^{p}(\theta),\rho\left(  \Gamma
^{q}(\theta),\ \cdot\ \right)  _{\theta}^{\mathcal{N}}\right)  _{\theta
}^{\mathcal{N}}. \label{49}%
\end{equation}

The definition of a generating functional of Green's functions $\mathsf{Z}%
\left(  (\partial_{\theta}\varphi^{\ast},\varphi^{\ast},\partial_{\theta
}\varphi,\mathcal{I})(\theta)\right)  \equiv\mathsf{Z}(\theta)$ as a path
integral, for a fixed $\theta$, is possible, within perturbation theory, by
introducing on $\mathcal{N}$ the Darboux coordinates, $\Gamma^{p}%
(\theta)=(\varphi^{a},\varphi_{a}^{\ast})(\theta)$, in a vicinity of solutions
of the equations ${\partial W(\theta)}/{\partial\Gamma^{p}(\theta)}=0$, so
that $\rho=1$ and $\omega^{pq}(\theta)=\mathrm{antidiag}(-\delta_{b}%
^{a},\delta_{b}^{a})$. The function
\begin{align}
\mathsf{Z}(\theta)=  &  \int d\mu\left(  \tilde{\Gamma}(\theta)\right)
\,d\Lambda(\theta)\exp\left\{  \left(  i/\hbar\right)  \left[  {W}\left(
\tilde{\Gamma}(\theta),\hbar\right)  +\left.  X\left(  \left(  \tilde{\varphi
},\tilde{\varphi}^{\ast}-{\varphi}^{\ast},\Lambda,\Lambda^{\ast}\right)
(\theta),\hbar\right)  \right|  _{\Lambda^{\ast}=0}\right.  \right.
\nonumber\\
&  -\left.  \left.  ((\partial_{\theta}\varphi_{a}^{\ast})\tilde{\varphi}%
{}^{a}+\tilde{\varphi}{}_{a}^{\ast}\partial_{\theta}^{r}{\varphi}{}%
^{a}-\mathcal{I}_{a}\Lambda^{a})(\theta)\right]  \right\}  \label{50}%
\end{align}
depends on an extended set of sources,%
\begin{align*}
&  (\partial_{\theta}\varphi_{a}^{\ast},\partial_{\theta}^{r}{\varphi}{}%
^{a},\mathcal{I}_{a})(\theta)=(-J_{a},\lambda^{a},I_{0a}+I_{1a}\theta),\\
&  (\vec{\varepsilon},\mathrm{gh})\partial_{\theta}\varphi_{a}^{\ast}%
=(\vec{\varepsilon},\mathrm{gh})\mathcal{I}_{a}+((1,0,1),1)=(\vec{\varepsilon
},-\mathrm{gh})\varphi^{a},
\end{align*}
to the superfields $(\varphi^{a},\varphi_{a}^{\ast},\Lambda^{a})(\theta)$,
where $\Lambda^{a}(\theta)=(\lambda_{0}^{a}+\lambda_{1}^{a}\theta)$ are
Lagrangian multipliers to independent non-Abelian hypergauges, see
\cite{BatalinTyutin},%
\begin{align*}
&  G_{a}(\Gamma(\theta)),a=1,...,\;k=n+\sum_{r=0}^{L}(2r+3)m_{r}%
,\;k=k_{+}+k_{-},\\
&  \mathrm{rank}\left\|  {\partial G_{a}(\theta)}/{\partial\Gamma^{p}(\theta
)}\right\|  _{{\partial W}/{\partial\Gamma}=G=0}=\overline{l},\;l=l_{+}%
+l_{-}=k.
\end{align*}
The functions $G_{a}(\Gamma(\theta))$, $(\vec{\varepsilon},\mathrm{gh}%
)G_{a}=(\vec{\varepsilon},\mathrm{gh})\mathcal{I}_{a}$, determine a boundary
condition for the gauge-fixing action, $X(\theta)=X\left(  (\Gamma
,\Lambda,\Lambda^{\ast})(\theta),\hbar\right)  $,%
\[
\left.  \partial_{r}X(\theta)/\partial\Lambda^{a}(\theta)\right|
_{\Lambda^{\ast}=\hbar=0}=G_{a}(\theta),
\]
defined on the direct sum $\mathcal{N}_{\mathrm{tot}}=\mathcal{N}\oplus\Pi
T^{\ast}\mathcal{K}$ of the manifolds $\mathcal{N}$ and $\Pi T^{\ast
}\mathcal{K}=\{(\Lambda^{a},\Lambda_{a}^{\ast})(\theta)\}$. Hypergauges in
involution, $(G_{a}(\theta),G_{b}(\theta))_{\theta}^{\mathcal{N}}=G_{c}%
(\theta)U_{ab}^{c}(\Gamma(\theta))$, obey different types of unimodularity
relations \cite{BatalinTyutin}, depending on a set of equations for which
$X(\theta)$ may be a solution, independently from ${W}(\theta)$, in terms of
the antibracket $(\,\cdot\,,\,\cdot\,)_{\theta}=(\,\cdot\,,\,\cdot\,)_{\theta
}^{\mathcal{N}}+(\,\cdot\,,\,\cdot\,)_{\theta}^{\mathcal{K}}$ and the operator
$\Delta(\theta)=(\Delta^{\mathcal{N}}+\Delta^{\mathcal{K}})(\theta)$,
trivially lifted from $\mathcal{N}$ to $\mathcal{N}_{\mathrm{tot}}$,%
\begin{equation}
1)\;(E(\theta),E(\theta))_{\theta}=0,\;\Delta(\theta)E(\theta)=0;\;2)\,\Delta
(\theta)\exp\left[  \frac{i}{\hbar}E(\theta)\right]  =0,\;E\in\{W,X\}.
\label{51}%
\end{equation}
The functions $G_{a}(\theta)$, assumed to be solvable with respect to
$\varphi_{a}^{\ast}(\theta)$, determine a Lagrangian surface, $\mathcal{Q}%
=\{({\varphi}^{\ast},\Lambda)(\theta)\}\subset\mathcal{N}_{\mathrm{tot}}$, on
which the restriction $\left.  X(\theta)\right|  _{\mathcal{Q}}$ is
non-degenerate. Given this, integration over $(\tilde{\varphi}^{\ast}%
,\Lambda)(\theta)$ in eq. (\ref{50}) determines a function, for $\partial
_{\theta}{\varphi}{}^{a}=\mathcal{I}_{a}=0$, whose restriction to the
Lagrangian surface $\{{\varphi}(\theta)\}\subset\mathcal{N}$ is also non-degenerate.

In view of the properties of $(W,X)(\theta)$, one can introduce an effective
action $\mathsf{\Gamma}(\theta)\equiv\mathsf{\Gamma}(\varphi,{\varphi}^{\ast
},\partial_{\theta}^{r}{\varphi},\mathcal{I})(\theta)$ defined, in the usual
manner, by means of a Legendre transformation of $\ln\mathsf{Z}(\theta)$ with
respect to $\partial_{\theta}{\varphi}_{a}^{\ast}(\theta)$,%
\begin{equation}
\mathsf{\Gamma}(\theta)=\frac{\hbar}{i}\ln\mathsf{Z}(\theta)+\left(
(\partial_{\theta}\varphi_{a}^{\ast})\varphi^{a}\right)  (\theta
),\;\varphi^{a}(\theta)=-\frac{\hbar}{i}\frac{\partial_{l}\ln\mathsf{Z}%
(\theta)}{\partial(\partial_{\theta}\varphi_{a}^{\ast}(\theta))}\,. \label{52}%
\end{equation}
The analysis of the properties of $(\mathsf{Z},\mathsf{\Gamma})(\theta)$ is
based on the following $\theta$-nonintegrable Hamiltonian-like system, which
contains an arbitrary $(\varepsilon_{P},\varepsilon)$-even $C^{\infty
}(\mathcal{N}_{\mathrm{tot}})$-function, $R(\theta)=R\left(  (\tilde{\Gamma
},\Lambda,\Lambda^{\ast})(\theta),\hbar\right)  $, with a vanishing ghost
number:%
\begin{equation}
\left\{
\begin{array}
[c]{l}%
\vspace{1ex}\partial_{\theta}^{r}\tilde{\Gamma}^{p}(\theta)=-i\hbar\left.
T^{-1}(\theta){\left(  \tilde{\Gamma}^{p}(\theta),T(\theta)R(\theta)\right)
_{\theta}}\right|  _{\Lambda^{\ast}=0}\,,\\
\vspace{1ex}\partial_{\theta}^{r}\Lambda^{a}(\theta)=-2i\hbar\left.
T^{-1}(\theta){\left(  \Lambda^{a}(\theta),T(\theta)R(\theta)\right)
_{\theta}}\right|  _{\Lambda^{\ast}=0}\,,\\
\partial_{\theta}^{r}\left(  \varphi_{a}^{\ast},\Lambda_{a}^{\ast}\right)
(\theta)=0\,,
\end{array}
\right.  \label{53}%
\end{equation}
where the function $T\left(  (\tilde{\Gamma},\Lambda,\Lambda^{\ast}%
)(\theta),\hbar\right)  \equiv T(\theta)$ has the form $T(\theta)=\exp\left[
\left(  i/\hbar\right)  (W-X)(\theta)\right]  $. Let us enumerate the
properties of $(\mathsf{Z},\mathsf{\Gamma})(\theta).$

1. The integrand in (\ref{50}) is invariant, for $\partial_{\theta}%
\varphi^{\ast}=\partial_{\theta}{\varphi}=\mathcal{I}=0$, with respect to the
\emph{superfield BRST transformations}%
\begin{equation}
\tilde{\Gamma}_{\mathrm{tot}}(\theta)=(\tilde{\Gamma},\Lambda,\Lambda^{\ast
})(\theta)\rightarrow\left(  \tilde{\Gamma}_{\mathrm{tot}}+\delta_{\mu}%
\tilde{\Gamma}_{\mathrm{tot}}\right)  (\theta),\;\delta_{\mu}\tilde{\Gamma
}_{\mathrm{tot}}(\theta)=\left.  \left(  \partial_{\theta}^{r}\tilde{\Gamma
}_{\mathrm{tot}}\right)  \right|  _{\check{\Gamma}_{\mathrm{tot}}}\mu,
\label{54}%
\end{equation}
having the form of a $\theta$-shift by a constant parameter $\mu$ along an
arbitrary solution $\check{\Gamma}_{\mathrm{tot}}(\theta)$ of the system
(\ref{53}), or, equivalently, along a vector field determined by the r.h.s. of
(\ref{53}), for $R(\theta)=1$. Here, the arguments of $(W,X)(\theta)$ are the
same as in definition (\ref{50}). The above statement can be verified with the
help of the identities%
\[
\left.  \partial_{r}X(\theta)/\partial F(\theta)\right|  _{\Lambda^{\ast}%
=0}=\partial_{r}(\left.  X(\theta)\right|  _{\Lambda^{\ast}=0})/\partial
F(\theta),\;F\in\{\Gamma^{p},\Lambda^{a}\}.
\]
Notice that the system (\ref{53}), for $R(\theta)=\mathrm{const}$, admits the
integral $(W+X)(\theta)$ in case $W$ and $X$ obey the first system in
(\ref{51}).

2. The vacuum function $\mathsf{Z}_{X}(\theta)\equiv\mathsf{Z}(0,\varphi
^{\ast},0,0)(\theta)$ is gauge-independent, namely, it does not change when
$X(\theta)$ is replaced by an $(X+\Delta X)(\theta)$ subject to the same
system in (\ref{51}) that holds for $X(\theta)$ and conforming to
nondegeneracy on the surface $\mathcal{Q}$. Indeed, this hypothesis implies
that the variation $\Delta X(\theta)$ obeys a system of linearized equations
with a nilpotent operator $Q_{j}(X)$, $j=1,2$,
\begin{equation}
Q_{j}(X)\Delta X(\theta)=0,\,\delta_{j1}\Delta(\theta)\Delta X(\theta
)=0;\;Q_{j}(X)=\mathrm{ad\,}X(\theta)-\delta_{j2}(i\hbar\Delta(\theta)),
\label{55}%
\end{equation}
where $j$ is identical to the number that labels that system in eqs.
(\ref{51}) for which $X(\theta)$ is a solution. Using the fact that solutions
$X(\theta)$ of every system in (\ref{51}) are proper, one can prove, by
analogy with the theorems of Ref. \cite{BatalinLavrovTyutin}, that the
cohomologies of the operator $Q_{j}(X)$ on the functions $f({\Gamma
}_{\mathrm{tot}}(\theta))\in C^{\infty}(\mathcal{N}_{\mathrm{tot}})$ vanishing
for ${\Gamma}_{\mathrm{tot}}(\theta)=0$ are trivial. Hence, the general
solution of eq. (\ref{55}) has the form
\begin{equation}
\Delta X(\theta)=Q_{j}(X)\Delta Y(\theta),\;\left(  \vec{\varepsilon
},\mathrm{gh},\frac{\partial}{\partial\theta}\right)  \Delta Y(\theta)=\left(
(1,0,1),-1,0\right)  ,\;\left.  \Delta Y(\theta)\right|  _{{\Gamma
}_{\mathrm{tot}}=0}=0, \label{56}%
\end{equation}
with a certain $\Delta Y\left(  \theta\right)  $. Now, making in
$\mathsf{Z}_{X+\Delta X}(\theta)$ a change of variables induced by a $\theta
$-shift by a constant $\mu$, related to the system (\ref{53}), and choosing%
\[
2R(\theta)\mu=\Delta Y(\theta),
\]
we find that $\mathsf{Z}_{X+\Delta X}(\theta)=\mathsf{Z}_{X}(\theta)$, and
conclude that the S-matrix is gauge-independent\footnote{Properties 1, 2 of
$\left.  \mathsf{Z}_{X}(\theta)\right|  _{\varphi^{\ast}=0}$ are valid for
arbitrary $\rho(\theta)$, $\Gamma^{p}(\theta)$ on the manifold $\mathcal{N}$.}
in view of the equivalence theorem \cite{KalloshTyutin}.

The above proof shows, due to (\ref{54}), that the system (\ref{53}) encodes
the BRST transformations for $R(\theta)=\mathrm{const}$, as well as continuous
anticanonical-like transformations in an infinitesimal form, with the scalar
fermionic generating function $R(\theta)\mu$, where $R(\theta)$ is arbitrary
and $\mu$ is constant.

Equivalently, following the ideas of Subsection 4.2, the above characteristics
of the generating functional of Green's functions can be derived from a
Hamiltonian-like system presented in terms of an even\ superfield Poisson
bracket in general coordinates (see footnote \thefootnote   ),
\begin{equation}
\left\{
\begin{array}
[c]{l}%
\vspace{1ex}\left.  \partial_{\theta}^{r}\tilde{\Gamma}^{p}(\theta)=-\left\{
\tilde{\Gamma}^{p}(\theta),Z^{W}[\tilde{\Gamma}]-(Z^{X}+i\hbar Z^{R}%
)[\tilde{\Gamma}_{\mathrm{tot}}]\right\}  \right|  _{\Lambda^{\ast}=0}\,,\\
\vspace{1ex}\left.  \partial_{\theta}^{r}\Lambda^{a}(\theta)=-2\left\{
\Lambda^{a}(\theta),Z^{W}[\tilde{\Gamma}]-(Z^{X}+i\hbar Z^{R})[\tilde{\Gamma
}_{\mathrm{tot}}]\right\}  \right|  _{\Lambda^{\ast}=0}\,,\\
\partial_{\theta}^{r}\left(  \varphi_{a}^{\ast},\Lambda_{a}^{\ast}\right)
(\theta)=0
\end{array}
\right.  \label{57}%
\end{equation}
with a linear combination of fermionic functionals related to the above
actions and a bosonic function by the rule
\begin{equation}
Z^{E}[{\Gamma}_{\mathrm{tot}}]=-\partial_{\theta}E({\Gamma}_{\mathrm{tot}%
}(\theta),\hbar),\;E\in\{W,X,R\}. \label{58}%
\end{equation}
If the actions $(W,X)(\theta)$ obey the first system in (\ref{51}), then the
functionals $Z^{W}$, $Z^{X}$, formally playing the role of the usual and
\emph{gauge-fixing} BRST charges, are nilpotent with respect to the even
Poisson bracket $\{\,\cdot\,,\,\cdot\,\}=\{\,\cdot\,,\,\cdot\,\}^{\Pi
T\mathcal{N}}+\{\,\cdot\,,\,\cdot\,\}^{\Pi T\mathcal{K}}$. Here, for instance,
the first bracket in the sum is defined on arbitrary functionals on ${\Pi
T\mathcal{N}}\times\{\theta\}$, via a $\theta$-local extension of the odd
bracket $(\,\cdot\,,\,\cdot\,)_{\theta}^{\Pi T\mathcal{N}}$ in (\ref{39}), as
follows:%
\begin{align}
&  \hspace{-1em}\left\{  F_{1},F_{2}\right\}  ^{\Pi T\mathcal{N}}\equiv\int
d\theta\frac{\delta_{r}F_{1}}{\delta\Gamma^{p}(\theta)}\omega^{pq}%
(\Gamma(\theta))\frac{\delta_{l}F_{2}}{\delta\Gamma^{q}(\theta)}%
=\partial_{\theta}(\mathcal{F}_{1}(\theta),\mathcal{F}_{2}(\theta))_{\theta
}^{\Pi T\mathcal{N}},\nonumber\\
&  \hspace{-1em}(\mathcal{F}_{1}(\theta),\mathcal{F}_{2}(\theta))_{\theta
}^{\Pi T\mathcal{N}}\equiv((\mathcal{L}_{p}^{r}\mathcal{F}_{1})\omega
^{pq}(\Gamma(\theta))\mathcal{L}_{q}^{l}\mathcal{F}_{2})(\theta
),\;F_{\mathrm{t}}[\Gamma]=\partial_{\theta}\mathcal{F}_{\mathrm{t}}%
((\Gamma,\partial_{\theta}\Gamma)(\theta),\theta), \label{59}%
\end{align}
where $\mathcal{L}_{q}^{l}(\theta)$ is the left-hand Euler--Lagrange
superfield derivative\footnote{The antibracket $(\,\cdot\,,\,\cdot\,)_{\theta
}^{\Pi T\mathcal{N}}$, identical, for $\mathcal{N}=\Pi T\mathcal{M}_{k}$, with
$(\,\cdot\,,\,\cdot\,)_{\theta}^{(\Gamma_{k},\partial_{\theta}\Gamma_{k})}$,
$k=\mathrm{ext}$, in (\ref{39}) lifts the operator $\Delta^{\mathcal{N}}$ in
(\ref{49}) to the nilpotent operator $\Delta^{\Pi T\mathcal{N}}$ acting in
$C^{\infty}\left(  \Pi T\mathcal{N}\times\{\theta\}\right)  $, defined exactly
as $\Delta^{\mathcal{N}}(\theta)$, although in terms of the antibracket
(\ref{59}).} with respect to $\Gamma^{q}(\theta)$.

Therefore, as in the case of the HS in (\ref{45}), we arrive at an
interpretation of BRST transformations, for a gauge theory with non-Abelian
hypergauges in Lagrangian formalism, in terms of the formal ``BRST charges''
$Z^{W}$, $Z^{X}$, as well as in terms of the functional $Z^{R}$ and the even
Poisson bracket\footnote{The construction of the latter bracket is different
from that of \cite{BatalinBeringDamgaard1}, where an odd superfield Poisson
bracket was derived from a $(t,\theta)$-local even bracket; however, it is
similar to the construction of Ref. \cite{GrigorievDamgaard}; see eqs. (27).}.
The system (\ref{57}) encodes the BRST transformations, for $Z^{R}=0$, as well
as the BRST and continuous canonical-like transformations with\thinspace
the\thinspace bosonic\thinspace generating functional\thinspace$Z^{R}\mu
$,\thinspace for\thinspace an\thinspace arbitrary\thinspace$Z^{R}$\thinspace
and\thinspace a\thinspace constant\thinspace$\mu$.

3. The functions $(\mathsf{Z},\mathsf{\Gamma})(\theta)$ obey the Ward
identities%
\begin{align}
&  \left\{  \left[  \partial_{\theta}\varphi_{a}^{\ast}(\theta)-\left(
\frac{\partial W}{\partial\tilde{\varphi}{}^{a}(\theta)}\right)  \left(
{i\hbar}\frac{\partial_{l}}{\partial(\partial_{\theta}\varphi^{\ast})}%
,{i\hbar}\frac{\partial_{r}}{\partial(\partial_{\theta}^{r}\varphi)}\right)
\right]  \frac{\partial_{l}}{\partial\varphi_{a}^{\ast}(\theta)}\right.
\nonumber\\
&  \left.  +\frac{i}{\hbar}\mathcal{I}_{a}(\theta)\frac{\partial_{l}}%
{\partial\Lambda_{a}^{\ast}(\theta)}\left.  X\left(  {i\hbar}\frac{\partial
_{l}}{\partial(\partial_{\theta}\varphi^{\ast})},{i\hbar}\frac{\partial_{r}%
}{\partial(\partial_{\theta}^{r}\varphi)}-\varphi^{\ast},\frac{\hbar}%
{i}\frac{\partial_{l}}{\partial\mathcal{I}},\Lambda^{\ast}\right)  \right|
_{\Lambda_{a}^{\ast}=0}\right\}  \mathsf{Z}(\theta)=0,\label{60}\\
&  \mathcal{I}_{a}(\theta)\frac{\partial_{l}}{\partial\Lambda_{a}^{\ast
}(\theta)}\left.  X\left(  {\varphi}^{b}+i\hbar(\mathsf{\Gamma}^{\prime\prime
}{}^{-1})^{bc}\frac{\partial_{l}}{\partial\varphi^{c}},{i\hbar}\frac{\partial
_{r}}{\partial(\partial_{\theta}^{r}\varphi)}-\frac{\partial_{r}%
\mathsf{\Gamma}}{\partial(\partial_{\theta}^{r}\varphi)}-\varphi^{\ast
},\frac{\partial_{l}\mathsf{\Gamma}}{\partial\mathcal{I}}+\frac{\hbar}%
{i}\frac{\partial_{l}}{\partial\mathcal{I}},\Lambda^{\ast}\right)  \right|
_{\Lambda_{a}^{\ast}=0}\nonumber\\
&  -\left[  \left(  \frac{\partial W}{\partial\tilde{\varphi}{}^{a}(\theta
)}\right)  \left(  {\varphi}^{b}+i\hbar(\mathsf{\Gamma}^{\prime\prime}{}%
^{-1})^{bc}\frac{\partial_{l}}{\partial\varphi^{c}},{i\hbar}\frac{\partial
_{r}}{\partial(\partial_{\theta}^{r}\varphi)}-\frac{\partial_{r}%
\mathsf{\Gamma}}{\partial(\partial_{\theta}^{r}\varphi)}\right)  \right]
\frac{\partial_{l}\mathsf{\Gamma}(\theta)}{\partial\varphi_{a}^{\ast}(\theta
)}+\frac{1}{2}\left(  \mathsf{\Gamma}(\theta),\mathsf{\Gamma}(\theta)\right)
_{\theta}^{(\Gamma)}=0, \label{61}%
\end{align}
with the notation $\mathsf{\Gamma}_{ab}^{\prime\prime}(\theta)\equiv
\frac{\partial_{l}}{\partial\varphi^{a}(\theta)}\frac{\partial_{r}}%
{\partial\varphi^{b}(\theta)}\mathsf{\Gamma}(\theta)\,$, $\mathsf{\Gamma}%
_{ac}^{\prime\prime}(\theta)(\mathsf{\Gamma}^{\prime\prime}{}^{-1}%
)^{cb}(\theta)=\delta_{a}\,^{b}$. In the symmetric form of these identities,
we have extended the standard set of sources $\partial_{\theta}\varphi
_{a}^{\ast}(\theta)$ used in the definition of the generating functional of
Green's functions in Abelian hypergauges.

The technique used in deriving the above identities is analogous to the
corresponding procedure of Refs. \cite{LO,LavrovOdintsovReshetnyak}, applied,
in the framework of the BV \cite{BV} and Batalin--Lavrov--Tyutin
\cite{BatalinLavrovTyutin} methods, to the problem of gauge dependence in
theories with composite fields. Thus, identities (\ref{60}) and (\ref{61})
follow from the corresponding system in (\ref{51}) for $(W,X)(\theta)$. For
instance, making the functional averaging of the second system in (\ref{51})
for $X(\theta)$,%
\begin{align}
&  \int d\Lambda(\theta)d\mu\left(  \tilde{\Gamma}(\theta)\right)  \exp\left[
\frac{i}{\hbar}\left(  {W}-(\partial_{\theta}\varphi_{a}^{\ast})\tilde
{\varphi}{}^{a}-\tilde{\varphi}{}_{a}^{\ast}\partial_{\theta}^{r}{\varphi}%
{}^{a}+\mathcal{I}_{a}\Lambda^{a}\right)  (\theta)\right] \nonumber\\
&  \times\left\{  \Delta(\theta)\exp\left[  \frac{i}{\hbar}X\left(
(\tilde{\varphi},\tilde{\varphi}^{\ast}-{\varphi}^{\ast},\Lambda,\Lambda
^{\ast})(\theta),\hbar\right)  \right]  \right\}  _{\Lambda^{\ast}=0}=0,
\label{62}%
\end{align}
and integrating by parts in (\ref{62}), with allowance for $(\partial
/\partial\tilde{\varphi}^{\ast}+\partial/\partial{\varphi}^{\ast})X(\theta
)=0$, we obtain identity (\ref{60}). Identities (\ref{60}) and (\ref{61}) take
the standard form in case $\partial_{\theta}{\varphi}^{a}=\mathcal{I}%
_{a}(\theta)=\theta=0$, which becomes more involved due to the quantities
$\left(  \partial_{\theta}W(\theta)/\partial\tilde{\varphi}{}^{a}%
(\theta)\right)  $, in the case of non-Abelian hypergauges.

In the special case of Abelian hypergauges, $G_{A}\left(  (\Phi,\Phi^{\ast
})(\theta)\right)  ={\Phi}_{A}^{\ast}(\theta)-\partial\Psi(\Phi(\theta
))/\partial\Phi^{A}(\theta)=0$, corresponding to the change of variables
(\ref{36}), for $(\varphi,\varphi^{\ast},W)=(\Phi,\Phi^{\ast},S_{\mathrm{H}%
;\mathrm{ext}})$, $\partial_{\theta}{\Phi}{}^{A}=\mathcal{I}_{A}=0$ (locally,
$\mathcal{N}=\Pi T^{\ast}\mathcal{M}_{\mathrm{ext}}$), the object
$\mathsf{Z}(\partial_{\theta}\Phi^{\ast},\Phi^{\ast})(\theta)$ takes the form
\begin{equation}
\mathsf{Z}\left(  \partial_{\theta}\Phi^{\ast},\Phi^{\ast}\right)
(\theta)=\int d\Phi(\theta)\mathrm{exp}\left\{  \frac{i}{\hbar}\left[
{S}_{\mathrm{H}}^{\Psi}\left(  {\Gamma}(\theta),\hbar\right)  -((\partial
_{\theta}\Phi_{A}^{\ast})\Phi^{A})(\theta)\right]  \right\}  . \label{63}%
\end{equation}
A $\theta$-local BRST transformation for $\mathsf{Z}\left(  \partial_{\theta
}\Phi^{\ast},\Phi^{\ast}\right)  (\theta)$ is given, for an HS defined on $\Pi
T^{\ast}\mathcal{M}_{\mathrm{ext}}$, with the Hamiltonian ${S}_{\mathrm{H}%
}^{\Psi}(\theta,\hbar)$ and a solution $\check{\Gamma}(\theta)$, by the change
of variables%
\begin{equation}
\Gamma^{p}(\theta)\rightarrow\Gamma^{(1){}p}(\theta)=\exp\left[  \mu
s^{l({\Psi})}(\theta)\right]  \Gamma^{p}(\theta),\;s^{l({\Psi})}(\theta
)\equiv\frac{\partial}{\partial\theta}-\mathrm{ad}{}S_{\mathrm{H}}^{\Psi
}(\theta,\hbar). \label{64}%
\end{equation}
Transformation (\ref{64}) with a constant $\mu$ is anticanonical, with
$\mathrm{Ber}\Vert\frac{\partial\Gamma^{(1)}(\theta)}{\partial\Gamma(\theta
)}\Vert=\mathrm{Ber}\Vert\frac{\partial\Phi^{(1)}(\theta)}{\partial\Phi
(\theta)}\Vert=1$, provided that $S_{\mathrm{H}}^{\Psi}(\theta,\hbar)$ is
subject to the first system in (\ref{51}).

The obvious permutation rule of the functional integral, $\varepsilon
(d\Phi(\theta))=0$,%
\[
\partial_{\theta}\int d\Phi(\theta)\mathcal{F}\left(  (\Phi,\Phi^{\ast
})(\theta),\theta\right)  =\int d\Phi(\theta)\left[  \frac{\partial}%
{\partial\theta}+(\partial_{\theta}V)(\theta)\right]  \mathcal{F}%
(\theta),\;(\partial_{\theta}V)(\theta)=\partial_{\theta}\Phi_{A}^{\ast
}(\theta)\frac{\partial}{\partial\Phi_{A}^{\ast}(\theta)}\,,
\]
yields, for $i\hbar\partial_{\theta}^{r}\ln\mathsf{Z}(\theta)=(\partial
_{\theta}\Phi_{A}^{\ast}\partial_{\theta}^{r}\Phi^{A})(\theta)-\partial
_{\theta}^{r}\mathsf{\Gamma}(\theta)$, the following relations:%
\begin{equation}
\left.  \partial_{\theta}\mathsf{Z}(\theta)\right|  _{\check{\Gamma}(\theta
)}=(\partial_{\theta}V)(\theta)\mathsf{Z}(\theta)=0,\;\left.  \partial
_{\theta}^{r}\mathsf{\Gamma}(\theta)\right|  _{\check{\Gamma}(\theta)}=\left(
\mathsf{\Gamma}(\Gamma(\theta)),\mathsf{\Gamma}(\Gamma(\theta))\right)
_{\theta}=0. \label{65}%
\end{equation}
When deriving eqs. (\ref{65}), we have taken into account the fact that the
functional averaging of the HS with respect to $\mathsf{Z}(\theta)$ and
$\mathsf{\Gamma}(\theta)$ has the form%
\begin{equation}
\left.  \langle\partial_{\theta}^{r}\Gamma^{p}\rangle\right|  _{\mathsf{Z}%
}=\left(  \frac{\hbar}{i}\mathsf{Z}^{-1}\frac{\partial\mathsf{Z}(\theta
)}{\partial\Phi_{A}^{\ast}(\theta)},-\partial_{\theta}\Phi_{A}^{\ast}%
(\theta)\right)  ,\;\langle\partial_{\theta}^{r}\Gamma^{p}\rangle=\left(
\langle\Gamma^{p}(\theta)\rangle,\mathsf{\Gamma}(\langle\Gamma(\theta
)\rangle)\right)  _{\theta}=\partial_{\theta}^{r}\langle\Gamma^{p}\rangle,
\label{66}%
\end{equation}
without the sign of average in (\ref{65}) for $\check{\Gamma}{}^{p}(\theta)$
and ${\Gamma}^{p}(\theta)$. Expressions (\ref{65}) relate the explicit form of
the Ward identities in a theory with Abelian hypergauges to the invariance of
the generating functional of Green's functions with respect to the superfield
BRST transformations.

\section{Connection between Lagrangian Quantizations}

The problem of establishing a correspondence between an LSM and a usual gauge
theory can be solved on the basis of a component form of the local
quantization in the following two ways: one is applicable to an arbitrary LSM,
another applies to theories of Yang--Mills type. This makes it possible to
establish a relation of the local superfield scheme with the known
formulations of Lagrangian quantization \cite{BV,BatalinTyutin}, as well as
with an extension (proposed below) of the superfield method \cite{LMR,GLM} to
the case of general coordinates.

\subsection{Component Formulation and its Relation to Batalin--Vilkovisky,
Batalin--Tyutin and Superfield Methods}

The objects of $\theta$-local quantization in the Lagrangian and Hamiltonian
formulations are related to the conventional description of a gauge theory by
means of a component representation of the variables $\Gamma_{\mathrm{CL}}%
^{P}$, $\Gamma_{k}^{p_{k}}$, $\Lambda^{a}$, $\mathcal{I}_{a}$, $\Gamma
_{k}^{p_{k}}(\theta)=\Gamma_{0k}^{p_{k}}+\Gamma_{1k}^{p_{k}}\theta$,
$k=\mathrm{tot}$, under the restriction $\theta=0$, for instance,
$(\mathcal{M},\mathcal{N}_{k},\Lambda^{a},\mathcal{I}_{a})\rightarrow
(\widetilde{\mathcal{M}},\left.  \mathcal{N}_{k}\right|  _{\theta=0}%
=\{\Gamma_{0k}^{p_{k}}\},\lambda_{0}^{a},I_{0a})$. Extracting a standard field
model from a classical description of a general gauge theory can be effected,
in addition to $\theta=0$, by various kinds of eliminating the functions
$\partial_{\theta}\mathcal{A}^{I}(\theta)$, $\mathcal{A}_{I}^{\ast}(\theta)$,
as well as the superfields $\mathcal{A}^{I}(\theta)$ that contain objects with
an incorrect spin-statistics relation, $\varepsilon_{P}(\mathcal{A}^{I})$
$\neq0$. A possible way of such elimination is provided by the conditions
$\mathrm{gh}(\mathcal{A}^{I})=-1-\mathrm{gh}(\mathcal{A}_{I}^{\ast})=0$,
$(\varepsilon_{P})_{I}=0$, and $\left(  \mathrm{gh},\partial/\partial
\theta\right)  S_{\mathrm{L}(\mathrm{H})}(\theta)=(0,0)$, mentioned in
Subsection 4.1. Another possibility is related to superfield BRST
transformations for theories of Yang--Mills type
\cite{Hull,NakanishiOjima,Malik}, in which a Lagrangian classical action
$S_{\mathrm{LYM}}(\theta)=S_{\mathrm{L}}\left(  \mathcal{A},\mathcal{D}%
_{\theta}\mathcal{A},\tilde{\mathcal{A}},\mathcal{D}_{\theta}\tilde
{\mathcal{A}}\right)  (\theta)$ is defined in terms of generalized Yang--Mills
superfields, $\mathcal{A}^{Bu}(z)$, $\mathcal{A}^{Bu}=(\mathcal{A}^{\mu
u},\mathcal{C}^{u})$, $u=1,...,r$, and matter superfields, $\tilde
{\mathcal{A}}(z)=(\Psi^{\varrho},\overline{\Psi}^{\sigma},\varphi^{f}%
,\varphi^{+g})(z)$, where $\Psi^{\varrho}$, $\overline{\Psi}^{\sigma}$,
$\varrho$, $\sigma=1,...,k_{1}$ , are spinor superfields, and $\varphi^{g}$,
$\varphi^{+h}$, $g$, $h=1,...,k_{2}$, are spinless ones. The superfields
$\mathcal{A}^{Bu}(z)$ and $\tilde{\mathcal{A}}(z)$ are defined on the
superspace $\mathcal{M}=\mathbb{R}^{1,3}\times\tilde{P}=\left\{  z^{B}%
=(x^{\mu},\theta)\right\}  $ and take values, respectively, in the adjoint and
vector representation spaces of an $r$-parametric Lie group. The action
$S_{\mathrm{LYM}}(\theta)$ can be written as%
\begin{equation}
S_{\mathrm{L{}YM}}(\theta)=\int d^{4}x\left[  \frac{1}{4}\mathcal{G}_{BC}%
{}^{u}\mathcal{G}^{CB{}u}(-1)^{\varepsilon_{B}}-i\overline{\Psi}^{\sigma
}\gamma^{B}\nabla_{B}{}_{\varrho}^{\sigma}\Psi^{\varrho}-\overline{\nabla}%
_{B}{}_{g}^{h}\varphi^{+g}\nabla^{B}{}_{f}^{h}\varphi^{f}+M(\tilde
{\mathcal{A}})\right]  (z), \label{67}%
\end{equation}
with an $\tilde{\mathcal{A}}(z)$-local gauge-invariant polynomial
$M(\tilde{\mathcal{A}})$, containing no derivatives over $z^{B}$. In
expression (\ref{67}), we have introduced the superfield strength
$\mathcal{G}_{BC}{}^{u}$=$i[\mathcal{D}_{B},\mathcal{D}_{C}]^{u}$%
=$\partial_{B}\mathcal{A}_{C}^{u}-(-1)^{\varepsilon_{B}\varepsilon_{C}%
}\partial_{C}\emph{A}_{B}^{u}+f^{uvw}\mathcal{A}_{B}^{v}\mathcal{A}_{C}^{w}$,
$\partial_{B}=(\partial_{\mu},\partial_{\theta})$ and the following covariant
derivatives, expressed through the matrix elements of the Hermitian generators
$\Gamma^{u}=\mathrm{diag}\left(  T^{u},\overline{T}^{u},\tau^{u}%
,\overline{\tau}^{u}\right)  $ of the corresponding Lie algebra:
\begin{equation}
\left(  \mathcal{D}_{B}^{uv},\nabla_{B}{}_{\varrho}^{\sigma},\nabla_{B}{}%
_{f}^{e},\overline{\nabla}_{B}{}_{h}^{g}\right)  =\partial_{B}\left(
\delta^{uv},\delta_{\varrho}^{\sigma},\delta_{f}^{e},\delta_{h}^{g}\right)
+\left(  f^{uwv},-i(T^{w})_{\varrho}^{\sigma},-i(\tau^{w})_{f}^{e}%
,-i(\overline{\tau}^{w})_{h}^{g}\right)  \mathcal{A}_{B}^{w}, \label{68}%
\end{equation}
where the coupling constant is absorbed into the completely antisymmetric
structure coefficients $f^{uvw}$. We have also used a generalization of
Dirac's matrices, $\gamma^{B}=(\gamma^{\mu},\gamma^{\theta})$, $\gamma
^{\theta}=(\gamma^{\theta})^{+}=\xi\mathbb{I}_{4}$, with a Grassmann scalar
$\xi$, $(\vec{\varepsilon},\mathrm{gh})\xi=((1,0,1),-1)$. The $\vec
{\varepsilon}$-grading and ghost number are nonvanishing for the superfields
$(\Psi,\overline{\Psi},\mathcal{C}^{u})$, namely, $\vec{\varepsilon}%
(\Psi,\overline{\Psi})=(0,1,1)$, $\vec{\varepsilon}(\mathcal{C}^{u})=(1,0,1)$,
$\mathrm{gh}(\mathcal{C}^{u})=1$. The functional $Z[\mathcal{A},\tilde
{\mathcal{A}}]=\partial_{\theta}S_{\mathrm{L{}YM}}(\theta)$ is invariant under
the infinitesimal general gauge transformations%
\begin{equation}
\delta_{g}\mathcal{A}^{I}(\theta)=\delta_{g}(\mathcal{A}^{Bu};\tilde
{\mathcal{A}})(z)=-\int d^{5}z_{0}\left(  \mathcal{D}^{Buv}%
(z)(-1)^{\varepsilon_{B}};i\Gamma^{v}\tilde{\mathcal{A}}(z)(-1)^{\varepsilon
(\tilde{\mathcal{A}})}\right)  \delta(z-z_{0})\xi^{v}(z_{0}), \label{69}%
\end{equation}
with arbitrary bosonic ($\vec{\varepsilon}_{\mathcal{A}_{0}}=\vec{0}$)
functions $\xi^{v}(z_{0})$ on $\mathcal{M}$, and with functionally-independent
generators $\hat{\mathcal{R}}{}_{\mathcal{A}_{0}}^{I}(\theta,\theta_{0}%
)\equiv\hat{\mathcal{R}}{}_{A_{0}}^{I}(\mathcal{A}(\theta),\theta,\theta_{0}%
)$. The condensed indices $I$, $\mathcal{A}_{0}$ of the theory in question,
$(I;\mathcal{A}_{0})=((B,u,\delta,\epsilon,f,h,x);(v,x_{0}))$, conform to the
relations, $\overline{N}>\overline{n}$, $\overline{M}=\overline{m}$,
$(\overline{m},\overline{M})=(\overline{m}_{0},\overline{M}_{0})$, provided
that%
\[
\overline{N}=(4r+2k_{2},r+8k_{1}),\ \overline{M}=(r,0),\ \overline
{n}=\overline{N}-(0,r),
\]
which holds for a reduced theory with the action\footnote{For $\theta=0$, the
functional $S_{\mathrm{YM}}(0)={S}_{0\mathrm{YM}}$ coincides with the
corresponding classical action of \cite{ChengLi}.} $S_{\mathrm{YM}}%
(\theta)=-S_{\mathrm{L{}YM}}\left(  \mathcal{A},0,\tilde{\mathcal{A}%
},0\right)  (\theta)$ on $\mathcal{M}_{\mathrm{cl}}=\{\mathcal{A}^{\mu
u},\tilde{\mathcal{A}}\}(z)$, in view of special \emph{horizontality
conditions} for the strength $\mathcal{G}_{BC}{}^{u}$ and certain subsidiary
conditions for the matter superfields $\tilde{\mathcal{A}}(z)$ in
\cite{Hull,NakanishiOjima},
\begin{equation}
\mathcal{G}_{BC}{}^{u}(z)=\mathcal{G}_{\mu\nu}{}^{u}(z),\;\left(
\nabla_{\theta}{}_{\eta}^{\delta}\Psi^{\eta},\overline{\nabla}_{\theta}%
{}_{\varrho}^{\sigma}\overline{\Psi}^{\varrho},\nabla_{\theta}{}_{e}%
^{f}\varphi^{e},\overline{\nabla}_{\theta}{}_{g}^{h}\overline{\varphi}%
^{g}\right)  (z)=(0,0,0,0). \label{70}%
\end{equation}
To extract a standard component model defined on $\left.  \mathcal{M}%
_{\mathrm{cl}}\right|  _{\theta=0}$ from a Hamiltonian LSM, it is sufficient
to eliminate, for $\theta=0$, the antifields $\mathcal{A}_{I}^{\ast}(\theta)$
of a theory of Yang--Mills type, by analogy with the prescription (\ref{70}),
i.e., by taking into account the relation between $\mathcal{A}_{I}^{\ast
}(\theta)$ and $\partial_{\theta}\mathcal{A}^{I}(\theta)$: see Section 3 and
the final remarks (item 1) of the Conclusion.

For the restricted LSM used in the Feynman rules of Section 4, the reduction
to the model of the multilevel formalism of Ref. \cite{BatalinTyutin} is
realized by the conditions%
\begin{equation}
\theta=0,\;\partial_{\theta}\varphi_{a}^{\ast}=\partial_{\theta}{\varphi}%
{}^{a}=\varphi_{a}^{\ast}=\mathcal{I}_{a}=0. \label{71}%
\end{equation}
In this case, the identification $(\rho,\omega^{pq})(\Gamma_{0})=(M,E^{pq}%
)(\Gamma_{0})$ implies the coincidence of $\left.  {(\,\cdot\,,\,\cdot
\,)_{\theta}}\right|  _{\theta=0}$ and $\Delta(0)$ with their counterparts of
\cite{BatalinTyutin}. Then the first-level functional integral $Z^{(1)}$ and
its symmetry transformations \cite{BatalinTyutin}
\begin{align*}
&  Z^{(1)}=\int d\lambda_{0}d\Gamma_{0}M(\Gamma_{0})\exp\left\{
\frac{i}{\hbar}\left(  W(\Gamma_{0})+G_{a}(\Gamma_{0})\lambda_{0}^{a}\right)
\right\}  ,\\
&  \left\{
\begin{array}
[c]{l}%
\delta\Gamma_{0}^{p}=(\Gamma_{0}^{p},-W+G_{a}\lambda_{0}^{a})\mu,\\
\delta\lambda_{0}^{a}=\left(  -U_{cb}^{a}\lambda_{0}^{b}\lambda_{0}%
^{c}(-1)^{\varepsilon_{c}}+2i\hbar V_{b}^{a}\lambda_{0}^{b}+2(i\hbar
)^{2}\tilde{G}^{a}\right)  \mu,
\end{array}
\right.
\end{align*}
coincide ($\lambda_{0}^{a}$ being replaced by the notation $\pi^{a}$ of
\cite{BatalinTyutin}), respectively, with $\left.  \mathsf{Z}_{X}(0)\right|
_{\varphi_{0}^{\ast}=0}$ and the BRST transformations $\delta_{\mu}%
\Gamma_{0\mathrm{tot}}$ (having the opposite signs) generated by the system
(\ref{53}) for $R(\theta)=1$. This coincidence is guaranteed by the choice of
$X(\theta)$ in the form%
\begin{equation}
X(\theta)=\left\{  G_{a}(\Gamma)\Lambda^{a}-\Lambda_{a}^{\ast}\left[
\frac{1}{2}U_{cb}^{a}(\Gamma)\Lambda^{b}\Lambda^{c}(-1)^{\varepsilon_{c}%
}-i\hbar V_{b}^{a}(\Gamma)\Lambda^{b}-(i\hbar)^{2}\tilde{G}^{a}(\Gamma
)\right]  \right\}  (\theta)+o(\Lambda^{\ast}), \label{72}%
\end{equation}
where $(V_{b}^{a},\tilde{G}^{a})(\theta)$, together with $(U_{cb}^{a}%
,G_{a})(\theta)$, define the unimodularity relations \cite{BatalinTyutin}. The
relation of the $\theta$-local quantization to the generating functional of
Green's function $Z[J,\phi^{\ast}]$ of the BV method \cite{BV} is obvious from
the identification $\mathsf{Z}\left(  \partial_{\theta}\Phi^{\ast},\Phi^{\ast
}\right)  (0)=Z[J,\phi^{\ast}]$ in (\ref{63}), where the action ${S}%
_{\mathrm{H}}^{\Psi}\left(  {\Gamma}_{0},\hbar\right)  $ of (\ref{36}) obeys
eq. (\ref{37}).

Another aspect of the restriction $\theta=0$ is that an arbitrary function
$\mathcal{F}(\theta)$ $=$ $\mathcal{F}\left(  (\Gamma,\partial_{\theta}%
\Gamma)(\theta),\theta\right)  \in C^{\infty}\left(  \Pi T\mathcal{N}%
\times\{\theta\}\right)  $ is represented by a functional $F[\Gamma]$ of the
superfield methods \cite{LMR,GLM} (in case $\Gamma^{p}=(\Phi^{A},\Phi
_{A}^{\ast})$, see the Introduction)%
\begin{equation}
F[\Gamma]=\int d\theta\theta\mathcal{F}(\theta)=\mathcal{F}\left(
\Gamma(0),\partial_{\theta}\Gamma,0\right)  \equiv\mathcal{F}(\Gamma
_{0},\Gamma_{1})\,. \label{73}%
\end{equation}
In the first place, formula (\ref{73}) implies the independence of $F[\Gamma]$
from $\partial_{\theta}^{r}\Gamma^{p}(\theta)=\Gamma_{1}^{p}$, in case
$F(\theta)=F(\Gamma(\theta),\theta)$. Secondly, formula (\ref{73}) is
fundamental in establishing a relation between the $\theta$-local antibracket
$(\,\cdot\,,\,\cdot\,)_{\theta}^{\mathcal{N}}$ and operator $\Delta
^{\mathcal{N}}(\theta)$, acting on $C^{\infty}\left(  N\times\{\theta
\}\right)  $, with a generalization to arbitrary $(\Gamma,\omega^{pq}%
,\rho)(\theta)$ of the flat functional operations $(\,\cdot\,,\,\cdot\,)$,
$\Delta$ of Refs. \cite{LMR, GLM}, identical to their counterparts of the BV
method in case $\Gamma^{p}=(\Phi^{A},\Phi_{A}^{\ast})$, $\omega^{pq}%
(\Gamma(\theta))=\mathrm{antidiag}\left(  -\delta_{B}^{A},\delta_{B}%
^{A}\right)  $, $\rho(\theta)=1$, and in case of a different odd Poisson
bivector, $\tilde{\omega}^{pq}(\Gamma(\theta),\theta^{\prime})=(1+\theta
^{\prime}\partial_{\theta})\omega^{pq}(\theta)$. The correspondence follows
from%
\begin{align}
&  \left.  \left(  {\mathcal{F}(\theta),\mathcal{G}(\theta)}\right)
{_{\theta}^{\mathcal{N}}}\right|  _{\theta=0}=\frac{\delta_{r}\mathcal{F}%
(\Gamma_{0})}{\delta\Gamma_{0}^{p}}\omega^{pq}(\Gamma_{0})\frac{\delta
_{l}\mathcal{G}(\Gamma_{0})}{\delta\Gamma_{0}^{q}}=\left(  F[\Gamma
],G[\Gamma]\right)  ^{\mathcal{N}},\nonumber\\
&  \left(  F[\Gamma],G[\Gamma]\right)  ^{\mathcal{N}}=\partial_{\theta}\left[
\frac{\delta_{r}F[\Gamma]}{\delta\Gamma^{p}(\theta)}\partial_{{\theta}%
^{\prime}}\left(  \tilde{\omega}^{pq}(\Gamma(\theta),\theta^{\prime
})\frac{\delta_{l}G[\Gamma]}{\delta\Gamma^{q}(\theta^{\prime})}\right)
\right]  (-1)^{\varepsilon(\Gamma^{p})+1},\label{74}\\
&  \left.  {\Delta^{\mathcal{N}}(\theta)\mathcal{F}(\theta)}\right|
_{\theta=0}=\Delta^{\mathcal{N}}(0)\mathcal{F}(\Gamma_{0})=\Delta
^{\mathcal{N}}F[\Gamma],\nonumber\\
&  \Delta^{\mathcal{N}}=\frac{1}{2}(-1)^{\varepsilon(\Gamma^{q})}%
\partial_{\theta}\partial_{\theta^{\prime}}\left[  \rho^{-1}[\Gamma
]\tilde{\omega}_{qp}(\theta^{\prime},\theta)\left(  \Gamma^{p}(\theta
),\rho\lbrack\Gamma]\left(  \Gamma^{q}(\theta^{\prime}),\cdot\right)
^{\mathcal{N}}\right)  ^{\mathcal{N}}\right]  , \label{75}%
\end{align}
where $\left(  \rho\lbrack\Gamma],\tilde{\omega}_{pq}(\theta^{\prime}%
,\theta)\right)  =\left(  \rho(\Gamma_{0}),\theta^{\prime}\theta{\omega}%
_{pq}(\theta)\right)  $ and
\[
\int d\theta^{\prime\prime}\tilde{\omega}^{pd}(\theta^{\prime},\theta
^{\prime\prime})\tilde{\omega}_{dq}(\theta^{\prime\prime},\theta)=\theta
\delta^{p}{}_{q}.
\]
When establishing the correspondence with the operations $(\,\cdot
\,,\,\cdot\,)$ and $\Delta$ of \cite{LMR, GLM} in (\ref{74}), (\ref{75}), we
have used a relation between the superfield and component derivatives:%
\[
{\delta_{l}}/{\delta\Gamma^{p}(\theta)}=(-1)^{\varepsilon(\Gamma^{p})}\left(
\theta{\delta_{l}}/{\delta\Gamma_{0}^{p}}-{\delta_{l}}/{\delta\Gamma_{1}^{p}%
}\right)  ,\;\Gamma_{1}^{p}=(\lambda^{A},-(-1)^{\varepsilon_{A}}J_{A}).
\]
In general coordinates, the action of the sum and difference $\partial
_{\theta}(V\pm U)^{\mathcal{N}}(0)$\textbf{\ }for $\mathcal{N}$ = $\left.  \Pi
T^{\ast}\mathcal{M}_{\mathrm{ext}}\right|  _{\theta=0}$ reduced to%
\[
\partial_{\theta}(V\pm U)(0)=\partial_{\theta}{\Phi}_{A}^{\ast}(\theta
)\partial/{\partial{\Phi}_{A}^{\ast}(0)}\pm\partial_{\theta}{\Phi}^{A}%
(\theta)\partial_{l}/{\partial{\Phi}^{A}(0)},
\]
is identical to the action of the generalized sum and difference of their
counterparts $V$, $U$ in \cite{LMR}:
\begin{align}
&  \left.  \left.  \partial_{\theta}(V-(-1)^{\mathrm{t}}U)^{\mathcal{N}%
}(\theta)\mathcal{F}(\theta)\right|  _{\theta=0}=\left(  {\mathcal{S}%
^{\mathrm{t}}(\theta),\mathcal{F}(\theta)}\right)  {_{\theta}^{\mathcal{N}}%
}\right|  _{\theta=0}\nonumber\\
&  =(V-(-1)^{\mathrm{t}}U)^{\mathcal{N}}F[\Gamma]=\left(  S^{\mathrm{t}%
}[\Gamma],F[\Gamma]\right)  ^{\mathcal{N}},\;\mathrm{t}=1,2,\nonumber\\
&  \mathcal{S}^{\mathrm{t}}(\theta)=(\partial_{\theta}\Gamma^{p})\omega
_{pq}^{\mathrm{t}}(\Gamma(\theta))\Gamma^{q}(\theta),\;S^{\mathrm{t}}%
[\Gamma]=\partial_{\theta}\left\{  \Gamma^{p}(\theta)\partial_{\theta^{\prime
}}\partial_{\theta}\left[  \tilde{\omega}_{pq}^{\mathrm{t}}(\theta
,\theta^{\prime})\Gamma^{q}(\theta^{\prime})\right]  \right\}  =\mathcal{S}%
^{\mathrm{t}}(0), \label{76}%
\end{align}
where the functions $\omega_{pq}^{\mathrm{t}}(\theta),\tilde{\omega}%
_{pq}^{\mathrm{t}}(\theta,\theta^{\prime})$, identical with $\omega
_{pq}(\theta)$ and $\tilde{\omega}_{pq}(\theta,\theta^{\prime})$ for
$\mathrm{t}=1$, are defined by%
\[
\tilde{\omega}_{pq}^{\mathrm{t}}(\theta,\theta^{\prime})=\theta\theta^{\prime
}\omega_{pq}^{\mathrm{t}}(\theta^{\prime})=-(-1)^{\mathrm{t}+\varepsilon
(\Gamma^{p})\varepsilon(\Gamma^{q})}\tilde{\omega}_{qp}^{\mathrm{t}}%
(\theta^{\prime},\theta),\ \omega_{pq}^{\mathrm{t}}(\theta)=(-1)^{\varepsilon
(\Gamma^{p})\varepsilon(\Gamma^{q})+\mathrm{t}}\omega_{qp}^{\mathrm{t}}%
(\theta)\,.
\]
The $\vec{\varepsilon}$-bosonic quantities $\mathcal{S}^{\mathrm{t}}(\theta) $
and $S^{\mathrm{t}}[\Gamma]$ with a vanishing ghost number play the role of
the symmetric Sp(2)-tensor $S_{ab}$ $(a,b=1,2)$ and anti-Hamiltonian $S_{0}$
of Ref. \cite{GLN}, which define (in terms of extended antibrackets) the
first-order operators of the modified triplectic algebra. In this case, the
additional functions ${\omega}_{pq}^{2}(\theta)$, $\tilde{\omega}_{pq}%
^{2}(\theta,\theta^{\prime})$ may be considered as quantities that define
another non-antisymplectic (non-Riemannian) nondegenerate structure on
$\mathcal{N}$. The $\theta$-local functional operators $\{\Delta^{\mathcal{N}%
},\partial_{\theta}V^{\mathcal{N}},\partial_{\theta}U^{\mathcal{N}}\}(\theta)$
anticommute for a fixed $\theta$,%
\begin{equation}
\lbrack E_{\mathrm{i}}^{\mathcal{N}}(\theta),E_{\mathrm{j}}^{\mathcal{N}%
}(\theta)]_{+}=0,\;\mathrm{i},\mathrm{j}=1,2,3,\;(E_{1},E_{2},E_{3}%
)=(\Delta\,,\partial_{\theta}V,\partial_{\theta}U), \label{77}%
\end{equation}
provided that $\mathcal{S}^{\mathrm{t}}(\theta)$ or $S^{\mathrm{t}}[\Gamma] $
is subject to%
\begin{equation}
\Delta^{\mathcal{N}}({\theta})\mathcal{S}^{\mathrm{t}}(\theta)=0,\;\left(
\mathcal{S}^{\mathrm{u}}(\theta),\mathcal{S}^{\mathrm{v}}(\theta)\right)
_{\theta}^{\mathcal{N}}=0,\;\mathrm{t},\mathrm{u},\mathrm{v}=1,2. \label{78}%
\end{equation}
Relations (\ref{78}), which hold, due to eqs. (\ref{74})--(\ref{77}), also for
functional objects (those without a $\theta$-dependence), follow from the
well-known properties of the antibracket (\emph{bilinearity},\emph{\ graded
antisymmetry},\emph{\ Leibniz rule},\emph{\ Jacobi identity}), and from the
rule of antibracket differentiation by the operator $\Delta^{\mathcal{N}%
}({\theta})$. The system (\ref{78}) determines the geometry of ${\mathcal{N}}$
by restricting the choice of both quantities $\omega_{pq}^{\mathrm{t}}%
(\theta)$, $\tilde{\omega}_{pq}^{\mathrm{t}}(\theta,\theta^{\prime})$. Notice
that a solution of eqs. (\ref{78}) always exists, for instance, $\omega
_{pq}^{\mathrm{t}}(\theta)=\mathrm{antidiag}\left(  \delta_{B}^{A}%
,(-1)^{\mathrm{t}}\delta_{B}^{A}\right)  $.

\subsection{Superfield Functional Quantization in General Coordinates}

Let us consider a generalization of the vacuum functional of the superfield
method \cite{LMR, GLM}, namely,%
\begin{equation}
Z_{X^{\prime}}^{\mathcal{N}}=\int d\mu\lbrack\Gamma]q^{\mathcal{N}}%
[\Gamma]\exp\left\{  \frac{i}{\hbar}\left(  W^{\prime}+X^{\prime}%
+\varkappa_{2}S^{2}\right)  [\Gamma]\right\}  , \label{79}%
\end{equation}
where $\varkappa_{2}$ is an arbitrary real number; $W^{\prime}$, $X^{\prime}$
are the quantum and gauge-fixing actions, defined on $\mathcal{N}$ and subject
to the equations
\begin{equation}
\frac{1}{2}(W^{\prime},W^{\prime})^{\mathcal{N}}+\mathit{V}W^{\prime}%
=i\hbar\Delta^{\mathcal{N}}W^{\prime},\;\frac{1}{2}(X^{\prime},X^{\prime
})^{\mathcal{N}}+\mathit{U}X^{\prime}=i\hbar\Delta^{\mathcal{N}}X^{\prime},
\label{80}%
\end{equation}
while the integration measure and the weight functional $q^{\mathcal{N}%
}[\Gamma]$ have the form%
\begin{equation}
d\mu\lbrack\Gamma]=\rho\lbrack\Gamma]\tilde{d}\Gamma,\;\tilde{d}\Gamma
={d}\Gamma_{0}{d}\Gamma_{1},\;q^{\mathcal{N}}[\Gamma]=\delta\left(  G_{a_{1}%
}^{\mathit{V}}(\Gamma(\theta))\right)  ,\,a_{1}=1,\ldots,\,\dim_{+}%
\mathcal{N}. \label{81}%
\end{equation}
In (\ref{80}), we have introduced a two-parameter set $\mathit{U}\left(
\varkappa_{1},\varkappa_{2}\right)  $, $\mathit{V}\left(  \varkappa
_{1},\varkappa_{2}\right)  $ of anti-commuting operators,%
\begin{equation}
\mathit{U}=\frac{1}{2}(-1)^{\mathrm{t}}\varkappa_{\mathrm{t}}(S^{\mathrm{t}%
}[\Gamma],\,\cdot\,)^{\mathcal{N}},\;\mathit{V}=\frac{1}{2}\varkappa
_{\mathrm{t}}(S^{\mathrm{t}}[\Gamma],\,\cdot\,)^{\mathcal{N}}, \label{82}%
\end{equation}
satisfying, together with $\Delta^{\mathcal{N}}$, the algebra (\ref{77}), for
arbitrary real numbers $\varkappa_{\mathrm{t}}$, whose choice admissible for
the existence of the functional integral fixes the form of $Z_{X^{\prime}%
}^{\mathcal{N}}$. This choice also fixes equations (\ref{80}), the admissible
boundary conditions for $W^{\prime}$, $X^{\prime}$, and the form of the
\emph{additional hypergauge conditions}, $G_{a_{1}}^{\mathit{V}}(\Gamma
(\theta))=0$, which are required to retain the explicit superfield form of the
vacuum functional. The independent functions $G_{a_{1}}^{\mathit{V}}%
(\Gamma(\theta))$ are equivalent to the set of functions $\mathit{V}\Gamma
^{p}(\theta)$: $G_{a_{1}}^{\mathit{V}}(\theta_{1})=\partial_{\theta}\left[
Y_{a_{1}p}(\Gamma(\theta_{1}),\theta)\mathit{V}\Gamma^{p}(\theta)\right]  $
with certain $Y_{a_{1}p}(\theta_{1},\theta)$ such that
\begin{equation}
\mathrm{rank}\left\|  P_{0}(\theta)\frac{\delta_{l}E_{\mathrm{t}}(\theta_{1}%
)}{\delta\Gamma^{q}(\theta)\ \ }\right\|  _{{\delta W^{\prime}/\delta
\Gamma=\delta X^{\prime}/\delta\Gamma}=G^{\mathit{V}}=0}=\left(
L_{\mathrm{t}}^{\mathit{V}},\dim_{+}\mathcal{N}-L_{\mathrm{t}}^{\mathit{V}%
}\right)  ,\,(E_{1},E_{2})=(G_{a_{1}}^{\mathit{V}},\mathit{V}\Gamma^{p}),
\label{83}%
\end{equation}
for some integers $L_{1}^{\mathit{V}}$, $L_{2}^{\mathit{V}}$: $0\leq
L_{1}^{\mathit{V}}$, $L_{2}^{\mathit{V}}\leq\dim_{+}\mathcal{N}$.

The basic properties of the functional $Z_{X^{\prime}}^{\mathcal{N}}$ are
analogous to properties 1, 2 of $\mathsf{Z}(\theta)$ in (\ref{50}), which are
encoded by a Hamiltonian-like system with an arbitrary functional $R[\Gamma]$,
$\left(  \vec{\varepsilon},\mathrm{gh}\right)  R=(\vec{0},0)$,%
\begin{equation}
\partial_{\theta}^{r}{\Gamma}^{p}(\theta)=\frac{\hbar}{i}T^{-1}[\Gamma]\left(
{\Gamma}^{p}(\theta),T[\Gamma]R\right)  ^{\mathcal{N}},\;T[\Gamma]=\exp\left[
\frac{i}{\hbar}\left(  W^{\prime}-X^{\prime}+\varkappa_{1}S^{1}\right)
\right]  . \label{84}%
\end{equation}
For instance, the \emph{superfield BRST transformations} $\delta_{\mu}{\Gamma
}^{p}(\theta)=\partial_{\theta}^{r}{\Gamma}^{p}(\theta)\mu$ for $Z_{X^{\prime
}}^{\mathcal{N}}$ are derived from (\ref{84}), with $R=1$, and from the
additional equations%
\begin{equation}
\left(  G_{a_{1}}^{\mathit{V}}(\Gamma(\theta)),W^{\prime}-X^{\prime}%
+\varkappa_{1}S^{1}\right)  ^{\mathcal{N}}=0\Longleftrightarrow\delta_{\mu
}G_{a_{1}}^{\mathit{V}}(\Gamma(\theta))=0, \label{85}%
\end{equation}
which ensure the BRST invariance of $q^{\mathcal{N}}$. In order to be valid
for any gauge theory with an admissible action, eqs.(\ref{85}) impose strong
restrictions on all the quantities $Y_{a_{1}p}(\theta_{1},\theta)$,
$\tilde{\omega}_{pq}^{\mathrm{t}}(\theta,\theta^{\prime})$, and consequently
on the geometry of $\mathcal{N}$. For example, the constant functions
$Y_{a_{1}p}(\theta_{1},\theta)$, $\tilde{\omega}_{pq}^{\mathrm{t}}%
(\theta,\theta^{\prime})$ belong to solutions of eqs. (\ref{85}). We, however,
do not restrict the consideration to this special case, assuming that eqs.
(\ref{85}) are fulfilled for any $W^{\prime}$, $X^{\prime}$. Some remarks are
in order concerning the status of the functional $q^{\mathcal{N}}[\Gamma]$.
Here, we do not consider the possibility of presenting the functions
$G_{a_{1}}^{\mathit{V}}(\theta)$ by an integral over new additional
superfields $\Lambda^{a_{1}}(\theta)$, following in part the prescription of
Ref. \cite{BatalinTyutin} that introduces so-called ``unimodularity involution
relations'' for $G_{a_{1}}^{\mathit{V}}(\theta)$ and modifies the BRST
transformations for the extended set of variables $(\Gamma^{p},\Lambda^{a_{1}%
})(\theta)$.

Choosing
\begin{equation}
\left(  \varkappa_{\mathrm{t}},\Gamma^{p},\rho,\tilde{\omega}_{pq}%
^{\mathrm{t}}(\theta,\theta^{\prime}),Y_{a_{1}p}(\theta_{1},\theta)\right)
=\left(  1,(\Phi^{A},\Phi_{A}^{\ast}),1,\theta\theta^{\prime}%
\mathrm{antidiag\,}(\delta_{B}^{A},(-1)^{\mathrm{t}}\delta_{B}^{A}%
),\delta(\theta_{1}-\theta)\delta_{Ap}\right)  , \label{86}%
\end{equation}
we obtain%
\begin{equation}
\left(  \mathit{V},\mathit{U},S^{2},q^{\mathcal{N}}\right)  =\left(
V,U,\partial_{\theta}(\Phi_{A}^{\ast}\Phi^{A})(\theta),\delta\left(
J_{A}\right)  \right)  , \label{87}%
\end{equation}
where $(V,U)=(-1)^{\varepsilon_{A}}\partial_{\theta}\left(  -\Phi_{A}^{\ast
}(\theta)\partial_{\theta}{\delta}/{\delta\Phi_{A}^{\ast}(\theta)},\Phi
^{A}(\theta)\partial_{\theta}{\delta_{l}}/{\delta\Phi^{A}(\theta)}\right)  $,
according to \cite{LMR}, and hence $Z_{X^{\prime}}^{\mathcal{N}}$, as well as
equations (\ref{80}), and BRST transformations, implied by (\ref{84}) for
$R=1$, coincide, respectively, with the vacuum functional $Z$,%
\[
Z=\int d\Phi d\Phi^{\ast}\delta(\partial_{\theta}\Phi^{\ast}(\theta
))\exp\left\{  \frac{i}{\hbar}\left(  W[\Phi,\Phi^{\ast}]+X[\Phi,\Phi^{\ast
}]+\partial_{\theta}(\Phi_{A}^{\ast}\Phi^{A})\right)  \right\}  ,
\]
with the equations $1/2(W,W)+VW=i\hbar\Delta W$, $1/2(X,X)-UX=i\hbar\Delta X$,
for $W=W^{\prime}$, $X=X^{\prime}$, and with the BRST symmetry transformations
\cite{GLM} for $Z$ (having the opposite signs in the r.h.s.)%
\[
\delta\Phi^{A}(\theta)=\mu U\Phi^{A}(\theta)+(\Phi^{A}(\theta),X-W)\mu
\,,\;\delta\Phi_{A}^{\ast}(\theta)=\mu V\Phi_{A}^{\ast}(\theta)+(\Phi
_{A}^{\ast}(\theta),X-W)\mu\,.
\]
In particular, choosing $X$ in terms of the gauge fermion $\Psi\lbrack
\Phi]=\Psi(\phi,\lambda)$, $X[\Phi,\Phi^{\ast}]=U\Psi\lbrack\Phi]$, first
realized in \cite{LMR}, we obtain the generating functional of Green's
functions $Z[\Phi^{\ast}]$ used in Section 1 in order to determine the
superfield effective action in Abelian hypergauges.

A complete correspondence between $Z_{X^{\prime}}^{\mathcal{N}}$ and the
functional $\left.  \mathsf{Z}_{X}(0)\right|  _{\varphi_{0}^{\ast}=0}$ in
(\ref{50}) can be established as follows: First, the functional $\varkappa
_{2}S^{2}$ is represented as $(1/2)(1+(-1)^{\mathrm{t}})\varkappa_{\mathrm{t}%
}S^{\mathrm{t}}$, so that the redefined actions%
\begin{equation}
W^{\prime\prime}=W^{\prime}+\frac{1}{2}\varkappa_{\mathrm{t}}S^{\mathrm{t}%
},\;X^{\prime\prime}=X^{\prime}+\frac{1}{2}(-1)^{\mathrm{t}}\varkappa
_{\mathrm{t}}S^{\mathrm{t}} \label{88}%
\end{equation}
obey eqs. (\ref{80}) without the operators $\mathit{V}$ and $\mathit{U}$.
Second, the actions $W(\theta)$ in (\ref{50}) and $W^{\prime\prime}[\Gamma]$,
as well as the quantities $\left.  X(\theta)\right|  _{\Lambda^{\ast}=0}$ in
(\ref{50}) and $X^{\prime\prime}[\Gamma]$, are related by formula (\ref{73}).
Third, the solvability of the hypergauges $G_{a}[\Gamma]$ with respect to the
fields $\varphi_{a}^{\ast}(\theta)$, on condition that $\Lambda^{a}%
(\theta)=\partial_{\theta}^{r}\varphi^{a}(\theta)$, implies, together with the
previous restriction, a linear dependence of $X^{\prime\prime}[\Gamma]$ on
$\Lambda^{a}(\theta)$ and its independence from $\partial_{\theta}\varphi
_{a}^{\ast}(\theta)$. Next, one should take into account the structure of the
generating equation for $X^{\prime\prime}[\Gamma]$, as well as the second
system in (\ref{51}) with (\ref{72}) for $X(\theta)$, and the fact that the
corresponding systems (\ref{84}), (\ref{53}), encoding the BRST
transformations, coincide with each other. The latter requires the
commutativity of $G_{a}[\Gamma]$ and the triviality of the unimodularity
relations, i.e., $\Delta^{\mathcal{N}}G_{a}=V_{b}^{a}=\tilde{G}^{a}=0$.
Finally, the measure $d\mu\lbrack\Gamma]q^{\mathcal{N}}$ in (\ref{79}) is made
identical to $\left.  d\mu(\Gamma(\theta))d\Lambda(\theta)\right|  _{\theta
=0}$ in (\ref{50}) by the choice $q^{\mathcal{N}}=\delta(\partial_{\theta
}\varphi^{\ast}(\theta))$. This choice can be realized by $(\varkappa
_{t},\tilde{\omega}_{pq}^{\mathrm{t}}(\theta,\theta^{\prime}),a_{1},Y_{a_{1}%
p}(\theta_{1},\theta))=\left(  1,\theta\theta^{\prime}\mathrm{antidiag}%
(\delta_{b}^{a},(-1)^{\mathrm{t}}\delta_{b}^{a}),a,\delta(\theta_{1}%
-\theta)\delta_{a_{1}p}\right)  $.

\section{Conclusion}

Let us summarize the main results of the present work:

We have proposed a $\theta$-local description of an arbitrary reducible
superfield theory as a natural extension of a usual gauge theory, defined on a
configuration space $\left.  {\mathcal{M}_{\mathrm{cl}}}\right|  _{\theta=0}$
of classical fields $A^{i}$, to a superfield model defined on extended
cotangent, $T_{\mathrm{odd}}^{\ast}\mathcal{M}_{\mathrm{CL}}\times\{\theta\}$,
and tangent, $T_{\mathrm{odd}}\mathcal{M}_{\mathrm{CL}}\times\{\theta\}$, odd
bundles, in respective Hamiltonian and Lagrangian formulations. It is shown
that the conservation under a $\theta$-evolution (defined by a Hamiltonian or
Lagrangian system providing a superfield extension of the usual extremals) of
a Hamiltonian action $S_{H}\left(  (\mathcal{A},\mathcal{A}^{\ast}%
)(\theta),\theta\right)  $, or, equivalently, of an odd counterpart of energy,
$S_{E}\left(  (\mathcal{A},\partial_{\theta}\mathcal{A})(\theta),\theta
\right)  $, is equivalent, in view of Noether's first theorem, to the validity
of a Hamiltonian or Lagrangian master equation, respectively.

Using non-Abelian hypergauges, we have constructed a $\theta$-local superfield
formulation of Lagrangian quantization for a reducible gauge model, extracted
from a general superfield model by the conditions of a manifest $\theta
$-independence of the classical action and the vanishing of ghost number and
auxiliary Grassmann parity (related to $\theta$) for the action and
$\mathcal{A}^{I}(\theta)$. In particular, we have proposed a new superfield
algorithm for constructing a first approximation to the quantum action in
powers of ghosts of the minimal sector, on the basis of interpreting the
reducibility relations as special gauge transformations of ghosts for an HS
with the Hamiltonian chosen as the quantum action. To investigate the
properties of BRST invariance and gauge-independence in a superfield form, for
the introduced generating functionals of Green's functions (including the
effective action), we have used \emph{two equivalent} Hamiltonian-like
systems. The first system is defined by a $\theta$-local antibracket, in terms
of a quantum action, a gauge-fixing action, and an arbitrary $\theta$-local
boson function, while the second (dual) system is defined by an even Poisson
bracket, in terms of fermion functionals corresponding to the above functions.
The two systems allow one to describe the BRST transformations and the
continuous (anti)canonical-like transformations in a manner analogous to the
relation between these transformations in the superfield Hamiltonian formalism
\cite{BatalinBeringDamgaard1}. We emphasize that, as a basis for the local
quantization, we have intensely used the first-level formalism of
\cite{BatalinTyutin}, whose main ingredient is the vacuum functional (however,
without recourse to the gauge-fixing action in a manifest form).

We have considered the problem of a \emph{dual description} for an $L$-stage
reducible gauge theory in terms of a BRST charge for a formal dynamical system
with first-class constraints of $(L+1)$-stage reducibility. It is shown that
this problem is a particular case of embedding a reducible special gauge
theory into a general gauge theory of the same stage of reducibility.

We have proposed an extension of functional superfield quantization
\cite{LMR,GLM} to the case of general antisymplectic manifold without
connection. It is shown that the condition of anti-commutativity for all
operators as well as the requirement of a correct transformation of the path
integral measure impose strong restrictions on the geometry of the manifold as
well as on \emph{additional hypergauge conditions} that determine the measure.

We have established the coincidence of the first-level functional integral
$Z^{(1)}$ in \cite{BatalinTyutin} with the local vacuum function of the
proposed quantization scheme, in case $\theta=0$ and $\varphi^{\ast}%
(\theta)=0$, $\left.  \mathsf{Z}_{X}(0)\right|  _{\varphi_{0}^{\ast}=0}$. A
correspondence is found between $\left.  \mathsf{Z}_{X}(0)\right|
_{\varphi_{0}^{\ast}=0}$ and the vacuum functional $Z_{X^{\prime}%
}^{\mathcal{N}}$ of the proposed extension of the superfield quantization
\cite{LMR,GLM}. It is shown that the above functionals coincide only in
Abelian hypergauges, with a trivial choice of the additional hypergauge conditions.

From the obtained results there follow the generating functional of Green's
functions and the effective action of the first-level formalism
\cite{BatalinTyutin}. It is observed that in case the quantum action
$W^{\prime}[\Gamma]$ depends on the superfields $\partial_{\theta}\Gamma
^{p}(\theta)$, or the gauge-fixing action $X^{\prime}[\Gamma]$ depends on the
same superfields more than linearly, the functional $Z_{X^{\prime}%
}^{\mathcal{N}}$ differs from $\left.  \mathsf{Z}_{X}(0)\right|  _{\varphi
_{0}^{\ast}=0}$ exactly as the functional $Z$ in \cite{GLM} differs from
$Z^{(1)}$ in \cite{BatalinTyutin}.

In connection with the discussed points, the following open problems seem to
be of interest:

1. One could obtain a Hamiltonian formulation of an LSM from a Lagrangian
formulation in the case of a degenerate Hessian supermatrix $(S_{\mathrm{L}%
}^{\prime\prime})_{IJ}(\theta)$ in (\ref{5}), and consider its relation to the
standard component description of a model. In this case, Dirac's algorithm in
terms of a $\theta$-local antibracket, under the conservation of primary
constraints in the course of $\theta$-evolution along a vector field defined
by an HS with a primary Hamiltonian in terms of antifields, would determine
all antisymplectic constraints for the classical superfields $\Gamma
_{\mathrm{CL}}^{P}(\theta)$. Among these constraints, there may exist a
subsystem of second-class ones, in the case of the degeneracy of the
supermatrix $\left\|  \mathcal{L}_{J}^{l}(\theta_{1})\left[  \mathcal{L}%
_{I}^{l}(\theta_{1})S_{\mathrm{L}}(\theta_{1})(-1)^{\varepsilon_{I}}\right]
\right\|  _{\Sigma}$ in (\ref{6}). It is interesting to apply the BFV method
to construct, in terms of a $\theta$-local Dirac's antibracket,\footnote{The
prescription for the first-level functional integral in terms of Dirac's
antibracket was considered in \cite{BatalinTyutin}} $(\,\cdot\,,\,\cdot
\,)_{\theta D}$, a triplet of $\theta$-local quantities $\tilde{S}%
_{\mathrm{H}}(\theta)$, $\tilde{\Omega}(\theta)$,$\tilde{\Psi}(\theta)$:
$(\varepsilon_{P},\varepsilon)$-even functions $\tilde{S}_{\mathrm{H}}(\theta)
$, $\tilde{\Omega}(\theta)$, commuting with respect to $(\,\cdot
\,,\,\cdot\,)_{\theta D}$ [by analogy with the Hamilton function and the
BFV--BRST charge in a $t$-local field theory] and an $(\varepsilon
_{P},\varepsilon)$-odd function $\tilde{\Psi}(\theta)$, which encodes the
dynamics of an LSM and its first-class constraint algebra, as well as fixes
the ``gauge'' arbitrariness in a space larger than $T_{\mathrm{odd}}^{\ast
}\mathcal{M}_{\mathrm{CL}}\times\{\theta\}$. In this connection, it seems
interesting to consider the question of how the construction of $\tilde
{S}_{\mathrm{H}}(\theta)$, $\tilde{\Omega}(\theta)$ and of the ``unitarizing
Hamiltonian'' $\tilde{\mathcal{S}}_{\mathrm{H}}(\theta)=\tilde{S}_{\mathrm{H}%
}(\theta)$ $+$ $(\tilde{\Omega}(\theta),\tilde{\Psi}(\theta))_{\theta D}$ is
related to the quantum action of the BV method.

2. From the solution of the dual problem of Subsection 4.2, found within the
classical description, there arise two natural questions: ``How does the
operator description of a formal dynamical system with a nilpotent BRST charge
and a quantum counterpart of the even Poisson bracket correspond to the
Lagrangian quantization of a gauge model?'' and ``Which ingredient of the
Lagrangian formulation should correspond to the formal supercommutator and the
Hilbert space of states?'' The mentioned problems seem to be related to the
correspondence \cite{BatalinMarnelius3} between Poisson brackets and their
operator counterparts of the opposite parity, as well as to the possibility
\cite{SharLyahov} of constructing a Lagrangian quantization procedure for more
general gauge theories that are determined, like higher-spin gauge fields
\cite{Vasiliev}, by non-Lagrangian equations of motion, $T_{i}(A)\neq
\delta\overline{S}(A)/\delta A^{i}$ for any $\vec{\varepsilon}$-bosonic
$\left.  \overline{S}(A)\in C^{\infty}\mathcal{M}_{\mathrm{cl}}\right|
_{\theta=0}$.

3. Notice that one of the possibilities of describing theories with
non-Abelian hypergauges within the superfield method \cite{LMR,GLM} consists
in enlarging the component spectrum of superfields $(\Phi^{A},\Phi_{A}^{\ast
})(\theta)$ by a Grassmann parameter ${\tilde{\theta}}$ unrelated to an
additional antiBRST symmetry. In this case, the inclusion of $(\Phi^{A}%
,\Phi_{A}^{\ast})(\theta)$ and the fields $\lambda_{A}^{\ast}$,
anticanonically conjugate to $\lambda^{A}$, into the general superfields
$(\Phi^{A},\overline{\Phi}_{A})(\theta,\tilde{\theta})$ is provided by the
relations%
\[
(\Phi^{A},\partial_{\tilde{\theta}}\overline{\Phi}_{A})(\theta,0)=(\Phi
^{A},\Phi_{A}^{\ast})(\theta),\ \overline{\Phi}_{A}(0,0)=\lambda_{A}^{\ast}.
\]

Finally, note that the procedure of $N=1$ local quantization has been recently
developed in \cite{Reshet2} as applied to the case of reducible general
hypergauges when independent hypergauge conditions cannot be determined in a
covariant manner on an antisymplectic manifold.

\textbf{Acknowledgments} The authors are grateful to P.M. Lavrov for useful
discussions. D.M.G. thanks the foundations FAPESP and CNPq for permanent
support. P.Yu.M. is grateful to FAPESP.

\end{document}